\documentclass[useAMS,usenatbib]{mn2e}
\usepackage{latexsym,graphicx,natbib}

\def\simless{{\th \rlap{\raise 0.5ex\hbox{$\scriptstyle  {<}$}}
    {\lower 0.3ex\hbox{$\scriptstyle  {\sim}$}} \th }}  
\def\simgreat{{\th \rlap{\raise 0.5ex\hbox{$\scriptstyle  {>}$}}
    {\lower 0.3ex\hbox{$\scriptstyle  {\sim}$}} \th }}  
\def\th{\thinspace}
\def\ts{{\raise 0.3ex\hbox{$\scriptstyle {\th \sim \th }$}}}
\def\kms{{\rm \th km\,s^{-1}}} 
\def\pct{{\rm pc^{-3}}} 
\def\masyr{\rm mas~{yr^{-1}}} 
\def\msun{M_\odot} 
\def\rsun{R_\odot}
\def\mtu{M_{\rm t1}}
\def\mtd{M_{\rm t2}}
\def\mpu{M_{\rm p1}}
\def\mpd{M_{\rm p2}}
\def\mtot{M_{\rm tot}}
\def\atar{a_{\rm t}}
\def\apro{a_{\rm p}}
\def\asec{{^{\prime\prime}}}
\def\aur{\rm AE~Aurig\ae}
\def\col{\rm \mu \th Columb\ae}
\def\ori{\rm \iota \th Orionis}
\def\aea{\rm AE~Aur}
\def\muc{\rm \mu~Col}
\def\ioo{\rm \iota~Ori}
\def\oria{\rm \iota~Ori~A}
\def\orib{\rm \iota~Ori~B}
\def\oriaa{\rm \iota~Ori~Aa} 
\def\oriaao{\rm \iota~Ori~Aa1} 
\def\oriaat{\rm \iota~Ori~Aa2} 
\def\oriab{\rm \iota~Ori~Ab}   
\def\emax{\mbox{$e_{\rm max}$}}

\title[N-body simulations of stars escaping from the Orion nebula]
      {N-body simulations of stars escaping from the Orion nebula}

\author[Gualandris, Portegies Zwart \& Eggleton] {
	Alessia Gualandris,$^{1,2}$\thanks{E-mail:
	alessiag@science.uva.nl} 
    	Simon Portegies Zwart$^{1,2}$
	and
	Peter P. Eggleton$^{3}$\\
$^1$    Astronomical Institute 'Anton Pannekoek',
	University of Amsterdam,
	the Netherlands \\
$^2$    Section Computational Science,
	University of Amsterdam,
	the Netherlands \\	
$^3$	Lawrence Livermore National Laboratory, 
	Livermore, CA\\
        }

\begin{document}

\date{Accepted by MNRAS.}

\pagerange{\pageref{firstpage}--\pageref{lastpage}} \pubyear{2003}

\maketitle

\label{firstpage}

\begin{abstract}
We study the dynamical interaction in which the two single runaway
stars $\aur$ and $\col$ and the binary $\ori$ acquired their unusually
high space velocity.  The two single runaways move in almost opposite
directions with a velocity greater than 100$\kms$ away from
the Trapezium cluster.
The star $\ori$ is an eccentric ($e\simeq$ 0.8) binary  
moving with a velocity of about 10$\kms$ at almost right angles 
with respect to the two single stars.  
The kinematic properties of the system suggest that a strong dynamical encounter 
occurred in the Trapezium cluster about $2.5\th$Myr ago.
Curiously enough, the two binary components have similar spectral type
but very different masses, indicating that their ages must be quite
different. This observation leads to the hypothesis that an exchange
interaction occurred in which an older star was swapped into the
original $\ori$ binary.
We test this hypothesis by a combination of numerical and theoretical
techniques, using N-body simulations to constrain the dynamical
encounter, binary evolution calculations to constrain the high orbital
eccentricity of $\ori$ and stellar evolution calculations to constrain
the age discrepancy of the two binary components.  
We find that an encounter between two low eccentricity ($0.4 \simless e \simless 0.6$) 
binaries with comparable binding energy, 
leading to an exchange and the ionization of the wider binary, 
provides a reasonable solution to this problem.  
\end{abstract}

\begin{keywords}
methods: N-body simulations - stars: binaries: spectroscopic - 
stars: individual: HD 34078, HD 37043, HD 38666
\end{keywords}

\section{Introduction}
\label{intro}
OB runaways are a subgroup of spectral type O and B stars which are
recognized by their unusually high velocity ($\simgreat 30\kms$), 
often away from known star forming regions (Blaauw, 1961).  
Stone (1979) also included stars at large distance (up to several kpc) 
from the Galactic plane for which no parent association could be found.
About 40 per cent of O stars and 5-10 per cent of B stars 
are runaways (Stone 1991), and only one in ten OB runaways 
has a known binary companion (Gies \& Bolton 1986).  

The peculiar velocities of about $50$ OB runaways have been determined
and the trajectories of $\ts$20 can be traced back 
to a nearby association (Hoogerwerf et al. 2000, 2001),
supporting the idea that these stars originated in stellar clusters 
from which they were later ejected with high velocities.

The two main mechanisms that have been proposed to explain 
the high velocities of OB runaways are:
\begin{itemize}
\item[1.] ejection upon a supernova in a binary system 
          (Zwicky 1957; Blaauw 1961);
\item[2.] ejection by a close multiple encounter in a star cluster
          (Poveda, Ruiz \& Allen, 1967).
\end{itemize}
We refer to scenario 1 and 2 as {\it supernova ejection} and
{\it dynamical ejection}, respectively.

In the supernova ejection scenario the OB runaway gets a high velocity
from the explosion of the companion star. If the binary is dissociated 
by the supernova, the companion star is ejected with a velocity 
of the order of its orbital velocity while if the binary remains bound, 
the mass loss in the supernova induces a recoil velocity 
in the center of mass of the binary (Blaauw 1961; Hills 1980; Tauris \& Takens 1998).  
This scenario predicts that about 30 per cent of the OB runaways should still
be in a binary with a compact companion.
Large natal kicks (up to 1000$\kms$) are observed in isolated pulsars 
(Lyne \& Lorimer 1994; Cordes \& Chernoff 1998), 
but are not sufficient to explain the low binarity fraction observed
for OB runaways.  
Even if the formation of a neutron star is accompanied by a kick drawn 
from a Maxwellian distribution with a mean of 600$\kms$, about
30 per cent of the binaries remain bound (Portegies Zwart 2000).
Selection effects may be important in preventing 
the observation of compact companions to the runaways (e.g. Colin et al. 1996), 
but the supernova ejection scenario seems insufficient to explain all
the known OB runaways (Portegies Zwart 2000).

The supernova ejection scenario, however, satisfactorily explains the
high velocities ($\ts50\kms$) of some known high-mass X-ray
binaries (van den Heuvel et al. 2000). In one of these, Vela X-1, the
high space velocity is inferred from the morphology of a bow shock
(Kaper et al. 1997) while the neutron star is visible as an X-ray pulsar.
The only known example of a binary dissociated by a natal kick is 
given by the O9.5V star $\zeta$ Oph and the pulsar PSR J1932+1059.
The large rotational velocity (Penny 1996) and helium abundance 
(Herrero et al. 1992) of $\zeta$ Oph indicate that a phase of mass transfer 
occurred, as expected for runaways ejected by a supernova explosion. 
Under this hypothesis, a natal kick of about 400$\kms$ can be inferred for the pulsar
(Hoogerwerf, de Bruijne \& de Zeeuw 2001).

An alternative scenario to the supernova ejection 
is the dynamical ejection scenario 
(Poveda, Ruiz \& Allen 1967; Gies \& Bolton 1986; Leonard 1990),
in which close encounters between binaries and/or single stars 
in dense stellar regions can lead to the ejection of stars with high velocity.
Young stellar clusters and OB associations in early evolutionary stages 
can have high stellar densities ($\ts10^5\th\pct$) in their cores
and therefore close encounters between single stars, 
binaries or higher order systems can be very frequent.
These encounters can result in the ejection of stars with high velocities.
O and B stars are more likely to participate in strong encounters
than low mass stars because of mass segregation and gravitational focusing.
As a consequence, the number of close encounters involving
O and B type stars can be high despite their short lifetime, 
especially when the cluster is young and the binary fraction is high. 

To help understanding the runaway mechanism we define 
the {\em kinematical age} $\tau_{\rm kin}$ 
as the time since the ejection from the parent association.
This time can be derived integrating the trajectory
of the runaway in the  Galactic potential.
At first order $\tau_{\rm kin}=d/v$, where $d$ is the distance traveled 
from the cluster and $v$ the velocity of the runaway. 
The kinematical age can help to identify the ejection mechanism.
A runaway which is ejected by a supernova explosion has a kinematical
age smaller than the age of the parent cluster because the primary star 
in the original binary evolved for a few million years before it exploded 
ejecting its companion.
Dynamical ejection is likely to occur when the cluster is very young
and therefore in this case the kinematical age is often similar 
to the age of the parent system.

An example of stars ejected by a dynamical encounter is provided by
the single runaways $\aur$ and $\col$ and the spectroscopic binary $\ori$,
whose trajectories in the Galactic potential have been traced back in time 
and intersect about $2.5\th$Myr ago at the expected location 
of the Trapezium cluster (Hoogerwerf et al. 2001).
The runaways $\aur$ and $\col$ show no evidence of binary evolution,
like a high rotational velocity or an increased helium abundance, 
and move in almost opposite directions away from the Trapezium cluster.
The kinematical and evolutionary properties of the system 
suggest the hypothesis that the stars were 
involved in a 4-body encounter that ejected them with high velocity.

In this paper we describe a combined use of direct N-body simulations
and stellar and binary evolution calculations to identify the type of
encounter that ejected $\aur$, $\col$ and $\ori$. 
The paper is organized as follows: In section $\S$ \ref{sec:prop} 
we describe the main observational properties of the stars and a possible
evolutionary model for $\ori$, $\aur$ and $\col$ 
and we propose a binary-binary encounter able to reproduce
the current positions, velocities and physical properties of the four stars; 
in section $\S$ \ref{sec:binbin} we describe the code used to perform
simulations of 4-body encounters and the classification of the outcomes;
in section $\S$ \ref{sec:simulations} we describe the choice of initial conditions
for the simulations, we present the main results obtained from the scattering
experiments and we compare them with the observed properties of the stars
while in section $\S$ \ref{sec:evolution} we present aspects of the evolution
of the $\ori$ binary after the encounter that can affect the model
for the dynamical encounter.

\section{The runaway stars AE Aurigae, $\mu$ Columbae and the runaway binary $\iota$ Orionis}
\label{sec:prop}

\subsection{The kinematics}
\label{sec:kin}
The single runaway stars $\aea$ and $\muc$ are O type stars 
moving away from the Orion star forming region.
Blaauw \& Morgan (1954) recognized them as a remarkable
pair of runaways having similar spectral types and moving in almost opposite 
directions  with space velocities of about 100$\kms$. 
At a distance of about $250\th$pc from the Trapezium cluster
the kinematical age of both stars is about $2.5\th$Myr.

The same kinematical age and parent cluster for the two stars
together with their relative straight angle of motion 
suggest a common origin in a dynamical encounter.
Gies \& Bolton (1986) advanced the hypothesis of a binary-binary 
interaction that ejected $\aur$ and $\col$ and left $\ori$ as the surviving
binary. They noticed that $\ioo$ is one of the most massive objects 
in the Orion nebula, has a high eccentric orbit 
and its component stars have relative orbital velocities similar 
to the space velocities of $\aea$ and $\muc$.

Parallaxes and proper motions provided by {\it Hipparcos}
together with radial velocity measurements (Turon et al. 1992) 
are shown in Table \ref{tab:vel} and allow precise determinations 
of the Galactic trajectories of the binary and the single stars.
The integration of the trajectories of $\aea$, $\muc$ and $\ioo$ 
(Hoogerwerf et al. 2001) in the Galactic potential shows that 
$2.5\th\pm\th0.2\th$Myr ago all the stars had a minimum separation of about 4$\th$pc.
The position and velocity of the parent cluster, derived 
from the orbit integration and the 4-body center of mass velocity, 
are consistent with those of the Trapezium cluster in the Orion nebula.

\begin{table}
  \caption{Parallax, distance from the sun and proper motions from the Hipparcos
    Catalogue (ESA 1997), radial velocities from the Hipparcos Input Catalogue
    (Turon et al. 1992), velocity relative to the Local Standard of Rest 
    and velocity relative to the center of mass of the 4-body system
    for $\aea$, $\muc$ and $\ioo$.}
  \label{tab:vel}
  \begin{center}
    \begin{tabular}{lccc}
      \hline
      & AE Aur & $\mu$ Col & $\iota$ Ori \\ 
      \hline
      \hline
      $\pi$ [mas]                 & 2.24 $\pm$ 0.74     & 2.52 $\pm$ 0.55    & 2.27 $\pm$ 0.65     \\
      d [pc]                      & $446_{-111}^{+220}$ & $397_{-71}^{+110}$ & $406_{-96}^{+185}$  \\
      $\mu_{\alpha^*}$ [$\masyr$] & -4.05 $\pm$ 0.66    & 3.01 $\pm$ 0.52    & 2.27 $\pm$ 0.65     \\
      $\mu_{\delta}$ [$\masyr$]   & 43.22 $\pm$ 0.44    & -22.62 $\pm$ 0.50  & -0.62 $\pm$ 0.47    \\
      $V_{\rm rad}$ [$\kms$]      & 57.5 $\pm$ 1.2      & 109.0 $\pm$ 2.5    & 28.7 $\pm$ 1.1      \\
      $V_{\rm LSR}$ [$\kms$]      & 111.9 $\pm$ 24.2    & 108.2 $\pm$ 4.2    & 17.4 $\pm$ 0.8      \\
      $V_{\rm cm}$ [$\kms$]       & 115 $\pm$ 5         & 103 $\pm$ 2        & 18 $\pm$ 1          \\
      \hline
    \end{tabular}
  \end{center}
\end{table}

We here summarize the arguments in favor of a common origin 
of the runaways $\aea$, $\muc$ and the $\ioo$ binary in the
Trapezium cluster:
\begin{itemize}
\item[1.] The trajectories of the single stars and the binary intersect
  2.5 Myr ago in the Trapezium cluster.
\item[2.] The velocity of the center of mass of the 4-body system is compatible 
  with the velocity of the Trapezium cluster.
\item[3.] The ages of all the four stars (see section \ref{sec:evol} and \ref{sec:age}) 
  are consistent with the range in ages of the stars in the cluster 
  (Palla \& Stahler 1999).
\item[4.] The high stellar density ($>2\times10^4\th\pct$, McCaughrean \& Stauffer 1994) 
  in the core of the Trapezium favors dynamical interactions.
\end{itemize}

\subsection{The stellar evolution}
\label{sec:evol}
The $\ori$ binary consists of a O9 III primary and a B0.8 III-IV secondary 
with a mass ratio of $q$=0.5 (Stickland et al. 1987) or $q$=0.57 (Marchenko 
et al. 2000).
Using the single-star evolutionary tracks by Schaller et al. (1992),
Bagnuolo et al. (2001) estimate a difference in age
of roughly a factor of two in the components and conclude
that the system did not co-evolve.
Therefore they propose that the system originated in a binary-binary encounter
in which an exchange interaction occurred.
The age difference cannot be explained by a phase of mass transfer
because of the high eccentricity of the binary: mass transfer would
have circularized the orbit of the binary.
Exchange encounters are known to alter the eccentricity of the 
interacting binaries and could provide a natural explanation 
for the high eccentricity of the system. 
Observational data on $\ioo$, $\aea$ and $\muc$ are presented
 in Table \ref{tab:prop}.

\begin{table}
  \caption{Properties for $\aur$, $\col$ (Gies and Lambert 1992;
    Bagnuolo, Riddle, Gies 2001) and the primary ($\oria$) 
    and secondary ($\orib$) component of $\ori$ 
    (Stickland et al. 1987; Bagnuolo, Gies and Penny 1994).}
  \label{tab:prop}
  \begin{center}
    \begin{tabular}{lcccc}
      \hline
      & AE Aur & $\mu$ Col & $\iota$ Ori A & $\iota$ Ori B \\ 
      \hline
      \hline
      Spectral type                  & O9.5 V & O9.5V/B0  & O9 III & B0.8 III-IV \\
      T$_{\rm eff}$ [$^\circ$K]      & 31420  & 31790  & 32500  & 27000  \\
      log(g)                         &  4.07  &  3.85  &  3.73  &  3.78  \\
      Mass [$\msun$]                 &    16  &    16  &    39  &    19  \\
      Radius [$\rsun$]               &   6.0  &   8.0  &    16  &    11  \\
      Period [days]                  &        &        & \multicolumn{2}{c}{29.134} \\
      Eccentricity                   &        &        & \multicolumn{2}{c}{0.764} \\
      \hline
    \end{tabular}
  \end{center}
\end{table}

Mason et al (1998) list $\ioo$ with a speckle companion at a separation 
of about 0.1 $\asec$. This could mean that $\ioo$ is the primary component 
of a hierarchical triple system. 
If the speckle companion is indeed bound to the binary, it has considerable 
consequences for the proposed origin of $\ioo$, as we discuss in $\S$ \ref{sec:speckle}. 
However, at the moment it is not clear whether the third component is 
bound to the binary. 

A possible evolutionary scheme for the system is shown in Table \ref{tab:scheme}.
Starting from the currently observed parameters reported in Table \ref{tab:prop},
the stellar and binary evolution package {\tt SeBa} (see Portegies Zwart \& Verbunt 1996) 
is used to infer masses and radii of each single star at the moment of the encounter.

\begin{table}
  \caption{Scheme of the evolutionary model adopted in this work 
    for $\ioo$, $\aea$ and $\muc$. Current values of mass, radius and age 
    for $\aea$, $\muc$ (Gies and Lambert 1992; Bagnuolo, Riddle, Gies 2001) 
    and the primary ($\oria$) and secondary ($\orib$) component of $\ioo$ 
    (Bagnuolo, Gies and Penny 1994)
    are used to predict masses and radii at the moment of the encounter.
    These parameters are used as initial conditions in the simulations.}
  \label{tab:scheme}
  \begin{center}
    \begin{tabular}{llcc}
      \hline
      T=0          &              \multicolumn{3}{l}{($\orib$, $\muc$)} \\
      \hline
      &  $\orib$   & 22   $\msun$ &  6.0 $\rsun$ \\
      &  $\muc$    & 18   $\msun$ &  5.0 $\rsun$ \\
      \hline
      T=3.5 Myr    &              \multicolumn{3}{l}{($\oria$, $\aea$)}  \\
      \hline
      &  $\oria$   &  42 $\msun$  &  8.5  $\rsun$ \\
      &  $\aea$    &  18 $\msun$  &  5.0  $\rsun$ \\
      \hline
      T=4.5 Myr    &              \multicolumn{3}{l}{($\oria$, $\aea$) + ($\orib$, $\muc$)} \\ 
      & \multicolumn{3}{c}{$\rightarrow$ ($\oria$, $\orib$) + $\aea$ + $\muc$} \\
      \hline
      &  $\oria$   &  42   $\msun$  &  10  $\rsun$ \\
      &  $\orib$   &  21.5 $\msun$  &  9.0 $\rsun$ \\
      &  $\aea$    &  18   $\msun$  &  5.5 $\rsun$ \\
      &  $\muc$    &  18   $\msun$  &  7.0 $\rsun$ \\
      \hline
      T=7 Myr      &              \multicolumn{3}{l}{($\oria$, $\orib$) + $\aea$ + $\muc$} \\
      \hline
      &  $\oria$   &  39 $\msun$   &  20  $\rsun$ \\
      &  $\orib$   &  20 $\msun$   &  16  $\rsun$ \\
      &  $\aea$    &  18 $\msun$   &  6.5 $\rsun$ \\
      &  $\muc$    &  18 $\msun$   &  10  $\rsun$ \\
      \hline
    \end{tabular}
  \end{center}
\end{table}

\subsection{The encounter}
\label{sec:enc}
\begin{figure}
  \caption{Example of a binary-binary encounter leading to the ionization of a binary
    and the exchange of one star. The two binaries are initially in the top right 
    and bottom left corner and approach in the direction indicated by the arrows.
    The binary containing $\muc$ is unbound and $\orib$ is exchanged in the other binary.
    As a result, the $\ori$ binary is formed and $\aur$ and $\col$ are ejected as singles.}
  \label{fig:exion}
  \begin{center}
    \leavevmode
    \includegraphics[width=8cm]{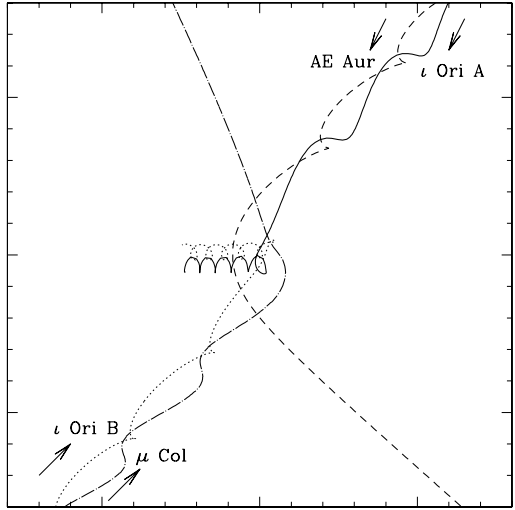}
  \end{center}
\end{figure}

We explore the hypothesis of a dynamical ejection for $\aea$, $\muc$ and $\ioo$
using numerical simulations of binary-binary encounters
and try to infer the interaction that is needed to produce 
the observed properties of the stars.  
First of all, one binary must be ionized during the encounter in order 
to eject the two single stars. In addition, an exchange is needed 
to solve the age discrepancy in the $\ori$ binary.
An example of a dynamical encounter producing a system like
$\ioo$, $\aea$ and $\muc$ is shown in Figure \ref{fig:exion}.
A binary consisting of $\orib$ and $\muc$ interacts with a binary
containing $\oria$ and $\aea$.
In the encounter the softer binary is unbound releasing $\muc$ and $\orib$,
which is exchanged in the other binary and together with $\oria$
forms the $\ori$ binary while $\aea$ and $\muc$ escape in almost opposite directions.

\section{Binary-binary scatterings}
\label{sec:binbin}

\subsection{The code and the setup of initial conditions}
\label{sec:code}
The numerical simulations described in this paper are carried out
with the {\tt scatter} package included in  the STARLAB software 
environment (McMillan \& Hut 1996; Portegies Zwart et al. 2001; 
{\tt http://www.ids.ias.edu/\~{}starlab}).
An N-body encounter is resolved integrating the equations of motions
of all the particles under the influence of their mutual gravitational forces
using a fourth-order Hermite predictor-corrector scheme (Makino \& Aarseth 1992).

We consider a {\it target} binary composed of stars of mass $\mtu$ and $\mtd$,
semi-major axis $\atar$ and eccentricity $e_{\rm t}$, 
and a {\it projectile} binary composed of stars of mass $\mpu$ and $\mpd$, 
semi-major axis $\apro$ and eccentricity $e_{\rm p}$.
Standard N-body units (Heggie \& Mathieu, 1986) are used in which 
$\mtu+\mtd$=1, $\atar$=1 and G=1. 
The relative velocity $v_{\infty}$ is given in units of the critical velocity $V_{\rm c}$
for which the total energy of the system 
in the 4-body center of mass reference frame is zero:
\begin{equation}
  V_{\rm c}=\sqrt{\frac{G}{\mu} \left(\frac{\mtu \mtd}{\atar}+\frac{\mpu \mpd}{\apro}\right)},
\end{equation}
where $\mu=(\mtu+\mtd)(\mpu+\mpd)/\mtot$ is the reduced mass and $\mtot$ 
the total mass of the system.
Additional parameters like the orbital phases and the relative orientation
of the two binaries are randomly drawn from uniform distributions
(Hut \& Bahcall 1983). The initial eccentricities are drawn from a thermal
distribution $P(e)=2e$ (Heggie 1975).

The impact parameter $b$ is chosen between zero and 
a maximum value according to an equal probability distribution for $b^2$.
The maximum value $b_{\rm max}$ is determined automatically for each experiment.
The code defines different cylindrical shells with the same cross-sectional area and performs 
an arbitrary number of scatterings in successive shells until all the encounters 
in the outermost shell result in a preservation of the binaries. 
The limiting impact parameter of the outermost shell defines
the value of $b_{\rm max}$ specific to the experiment under consideration. 

Energy conservation in our simulations is typically better than 
one part in $10^6$. 
The error in energy conservation is checked in each experiment 
and if it exceeds $10^{-5}$ the encounter is rejected.
The accuracy in the code is chosen in such a way that at most
a few per cent of the encounters are rejected.

\subsection{Classification system for binary-binary encounters}
\label{sec:outcomes}
When the integration of the equations of motion is stopped 
(see appendix \ref{app:stop} for the stopping criteria)
the encounters are classified.
During the calculation we keep track of the nearest neighbors of
all the stars and we compute the distances between them.
If two stars approach each other within a distance smaller than the sum of
their radii we classify the encounter as a collision. 

Many outcomes are possible in binary-binary interactions:
two bound systems can be left, or one binary and two single stars, 
or a triple and a single star, or four single stars. 
The encounters are classified as follows:
(i) {\it preservation} or {\it flyby} if the two original binaries remain bound,
even though strongly perturbed;
(ii) {\it exchange} if one star is exchanged by the binaries and
two new binaries are formed;
(iii) {\it ionization} if one binary is unbound leaving a binary 
and two single stars recoiling to infinity;
(iv) {\it exchange-ionization} if one binary is unbound and one star is exchanged
in the other binary;
(v) {\it triple} if one binary is unbound and one of the stars is captured 
by the other binary in a bound stable orbit while the other star is ejected as single;
(vi) {\it total ionization} if both binaries are unbound and four single stars 
recede to infinity.

\begin{figure*}
  \caption{Examples of different outcomes in binary-binary scattering events 
    and the corresponding Feynman diagrams. In the diagrams the paths of two stars are placed 
    close together when they are bound,
    and the wavy lines represent gravitational perturbations.}
  \label{fig:orbits}
  \includegraphics[width=5.5cm]{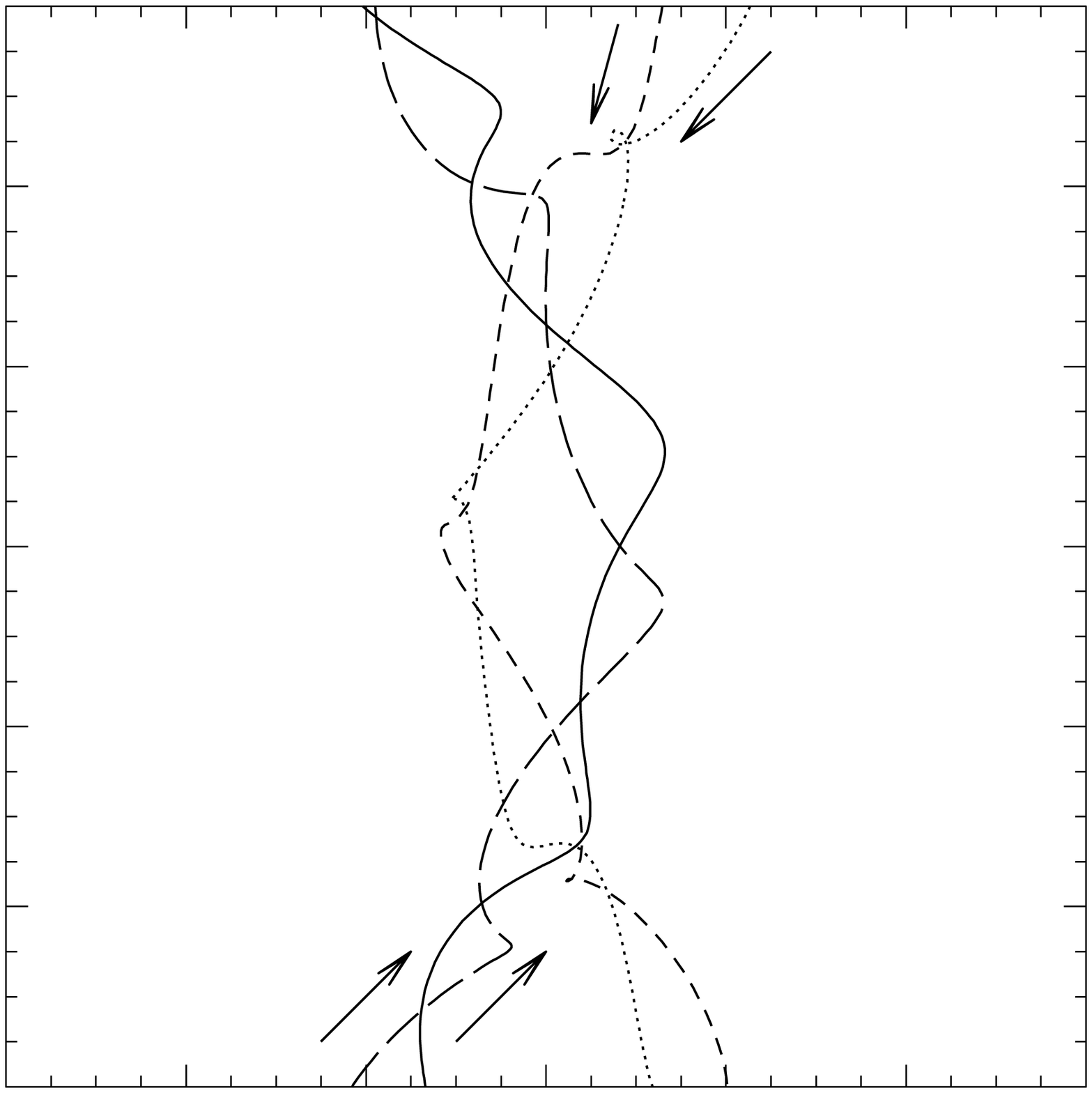}
  \includegraphics[width=5.5cm]{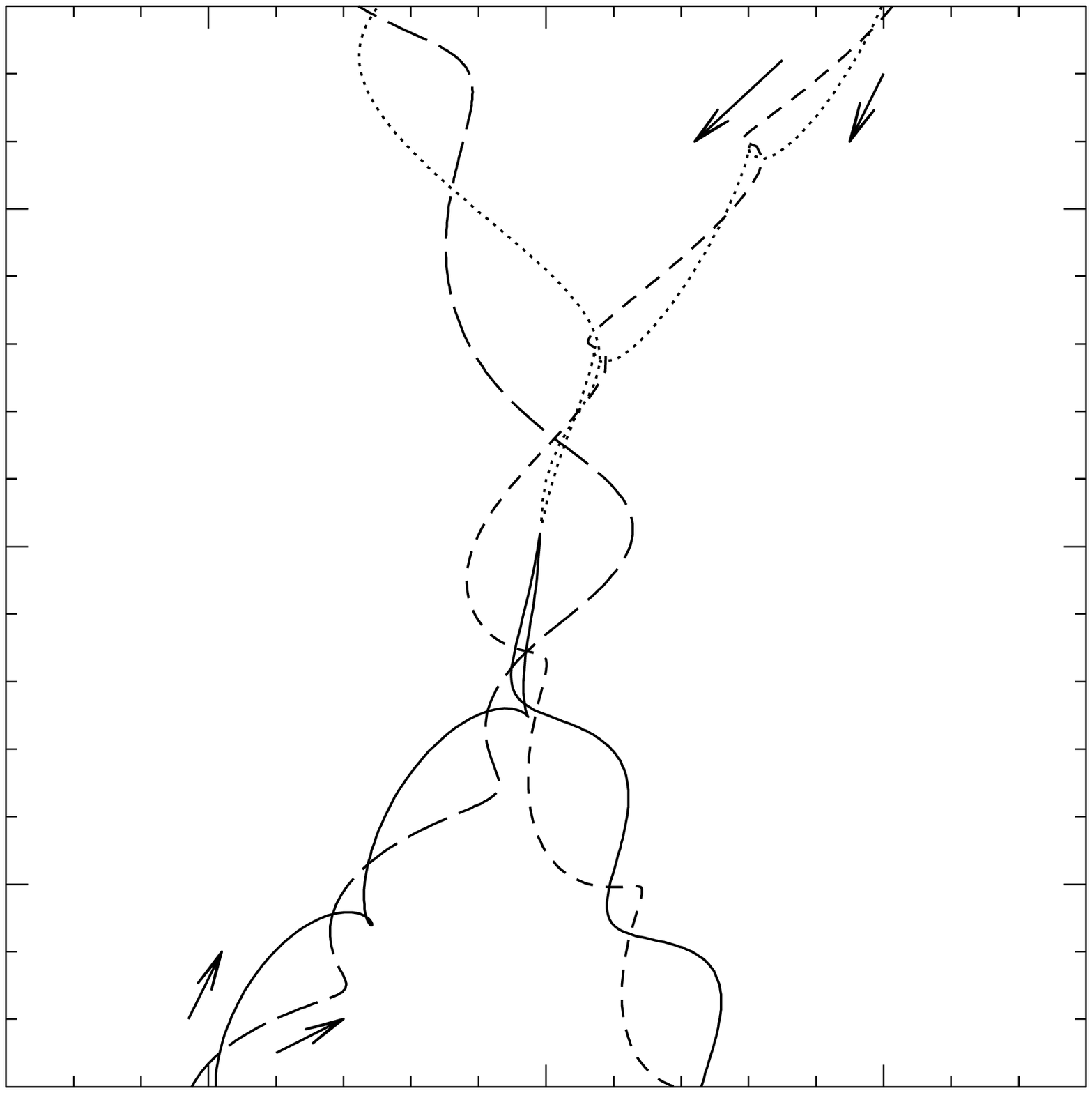}
  \includegraphics[width=5.5cm]{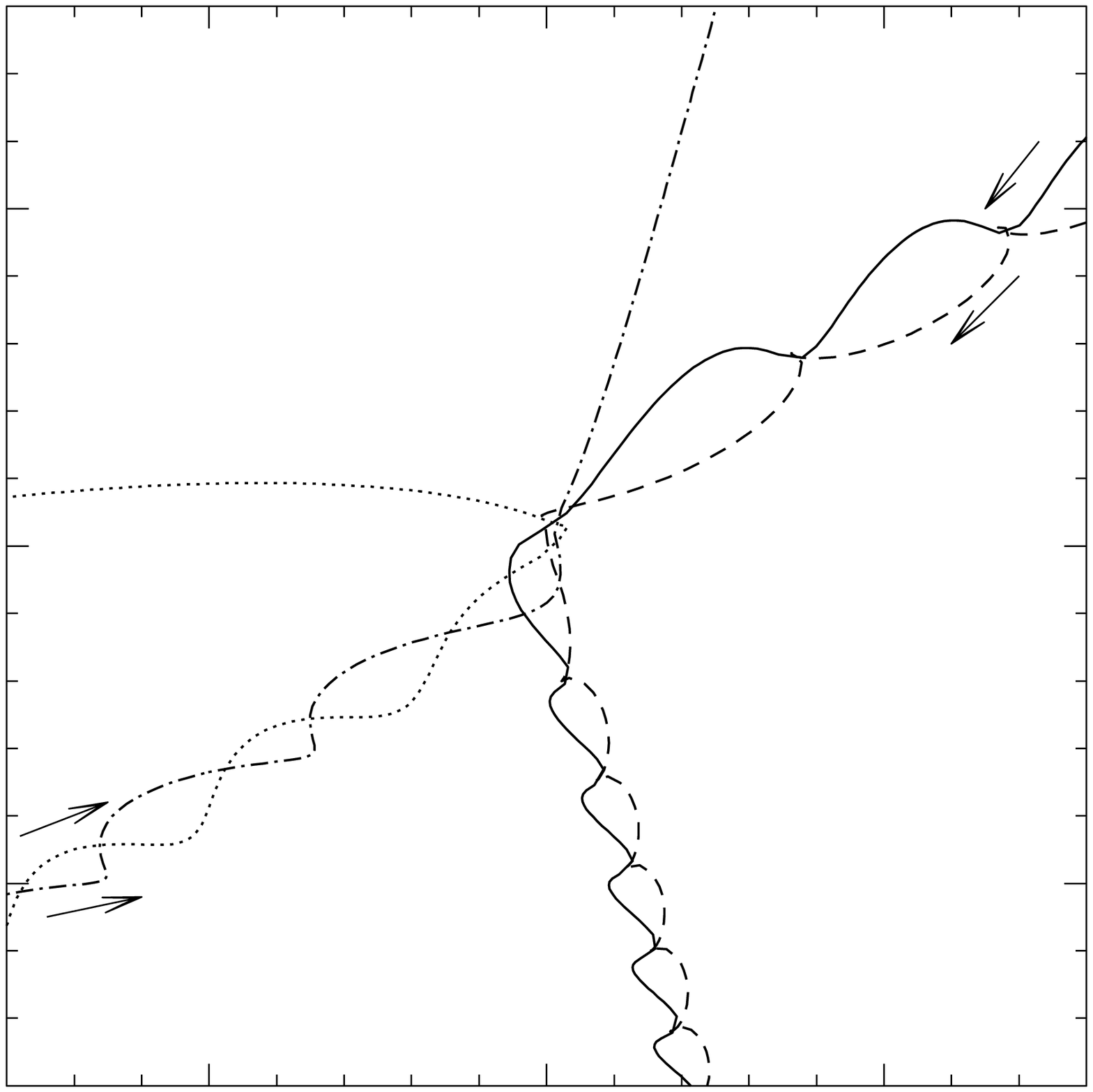}
  \includegraphics[width=5.5cm]{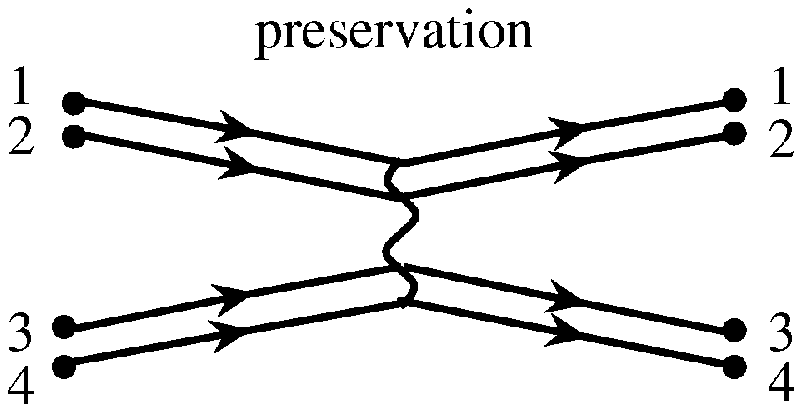}
  \includegraphics[width=5.5cm]{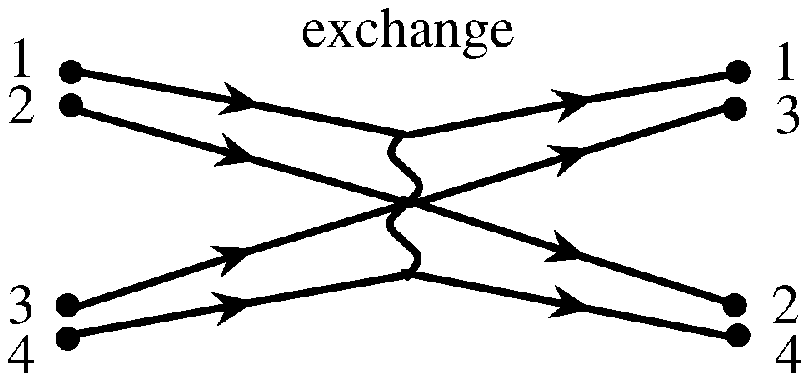}
  \includegraphics[width=5.5cm]{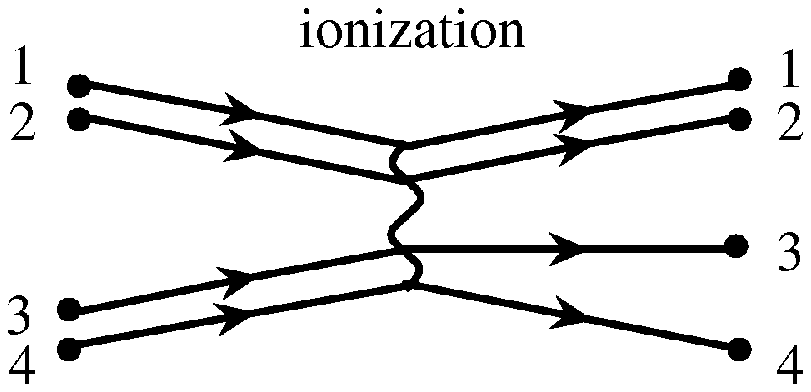}
  \includegraphics[width=5.5cm]{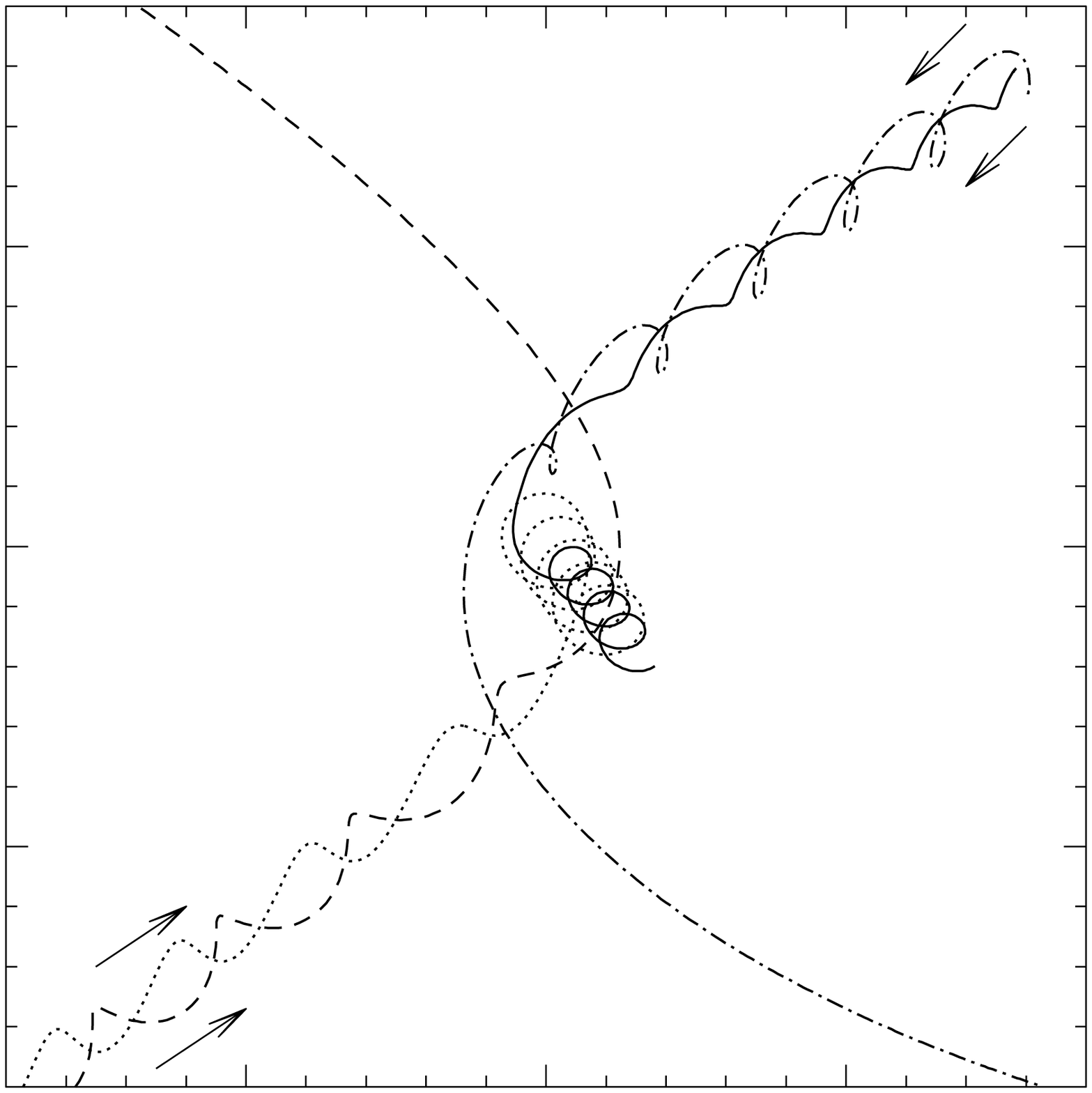}
  \includegraphics[width=5.5cm]{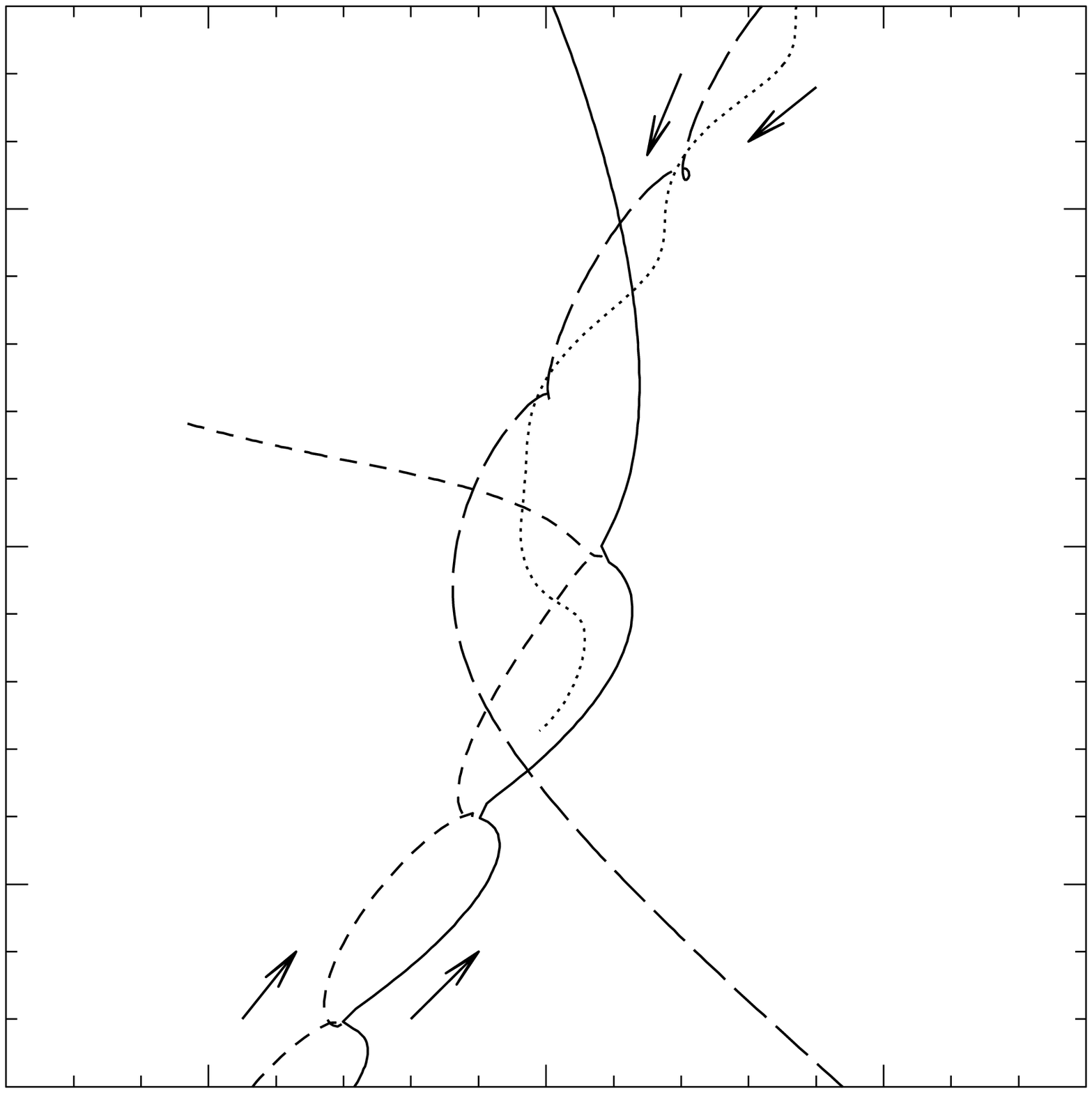}
  \includegraphics[width=5.5cm]{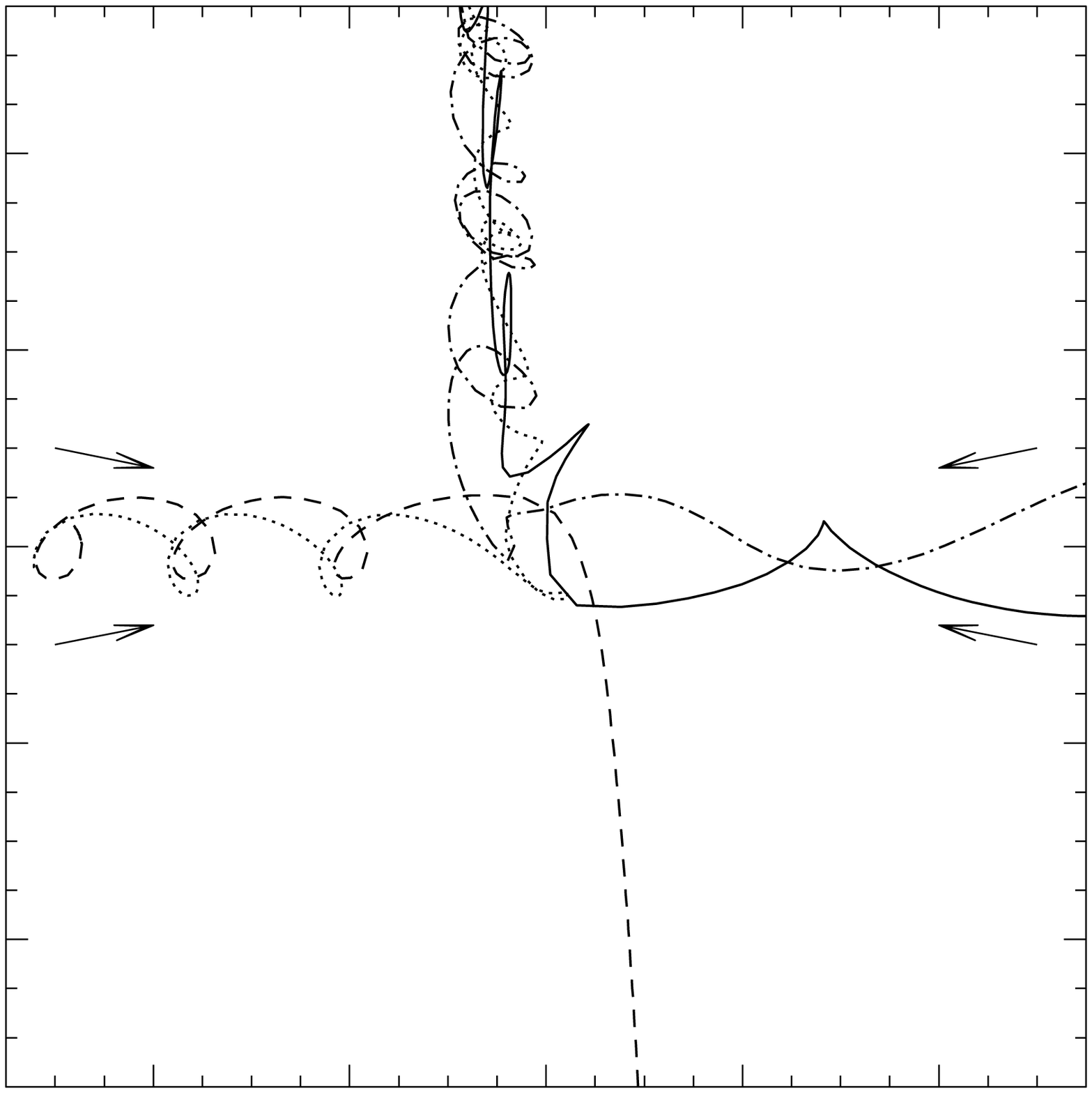}
  \includegraphics[width=5.5cm]{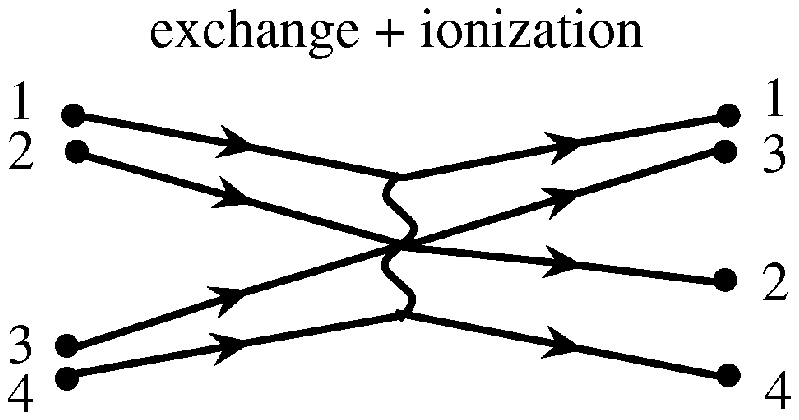}
  \includegraphics[width=5.5cm]{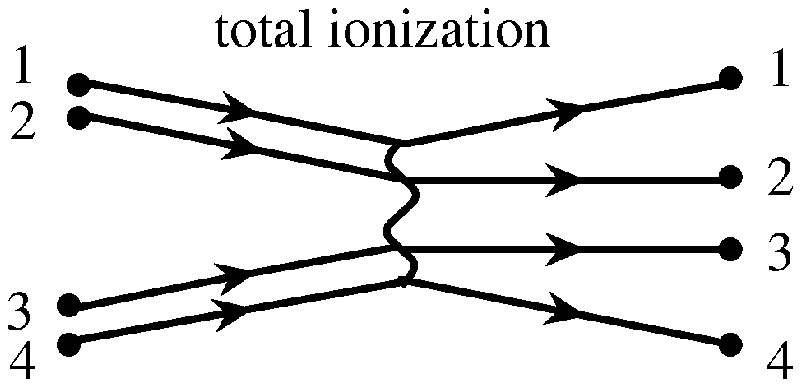}
  \includegraphics[width=5.5cm]{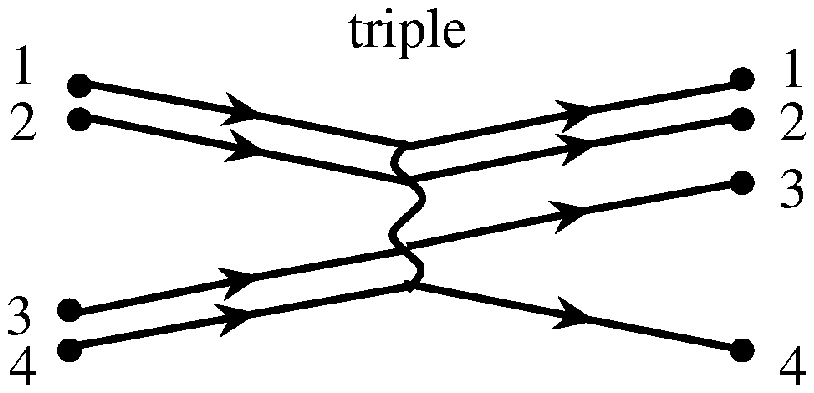}
\end{figure*}

In Figure \ref{fig:orbits} we present different possible outcomes 
(and corresponding Feynman diagrams) of binary-binary interactions.

\subsection{Testing the code}
\label{sec:test}
In order to test our scattering package, we performed a series
of binary-binary scattering experiments and compared
the results with those of Mikkola (1983).
He used a hybrid code containing three different integration
methods: the hierarchy method, where one or more of the relative motions
in the system are integrated as a perturbed keplerian orbit,
direct integration and Heggie's general N-body regularization method
(Heggie 1974).
Mikkola considered two identical binaries with equal mass stars
approaching each other with various relative velocities.
We classify his results in the following way:
{\em unresolved} for an encounter that cannot be classified as any 
of the other types at the moment the integration is stopped; 
{\em flyby} if two binaries are in a hyperbolic relative orbit;
{\em stable triple} for a hierarchical triple and one escaping star;
{\em single ionization} if one binary and two escaping stars remain
and finally {\em total ionization} if both binaries are disrupted.
The frequencies of the different outcomes are reported in Table \ref{tab:mikk} 
as a function of the relative kinetic energy $T_{\infty}$ between the binaries
together with the results by Mikkola. 
The agreement is in most cases better than the 1-$\sigma$ level
and always within 2-$\sigma$. 

\begin{table*}
  \caption{Comparison of the frequencies of different types of outcomes 
    with the results by Mikkola (1983). For the scattering experiments 
    we consider binaries with four equal-mass stars and equal semi-major axes.}
  \label{tab:mikk}
  \begin{tabular}{cccccccccccc}
    \hline
    \hline
    $T_{\infty}$  & $v_{\infty}$ & \multicolumn{2}{c}{unresolved} & \multicolumn{2}{c}{flyby} 
    & \multicolumn{2}{c}{stable triple} & \multicolumn{2}{c}{single ionization} 
    & \multicolumn{2}{c}{total ionization} \\
    Mikkola & Starlab & Mikkola & Starlab & Mikkola & Starlab & Mikkola & Starlab & 
    Mikkola & Starlab & Mikkola & Starlab \\
    \hline
    \hline
    0.10 & 0.316 & 7 & 0 & 358 & 350 & 38 & 32 & 97 & 118  & 0 & 0 \\
    0.25 & 0.5   & 6 & 0 & 387 & 409 & 28 & 20 & 79 &  71  & 0 & 0 \\
    0.50 & 0.707 & 0 & 0 & 430 & 436 &  5 & 10 & 65 &  54  & 0 & 0 \\
    0.75 & 0.866 & 2 & 0 & 452 & 451 &  2 &  1 & 44 &  48  & 0 & 0 \\
    1.00 & 1.0   & 1 & 0 & 450 & 442 &  0 &  0 & 49 &  58  & 0 & 0 \\
    1.50 & 1.225 & 0 & 0 & 466 & 452 &  0 &  0 & 31 &  45  & 3 & 3 \\
    \hline
  \end{tabular}
\end{table*}

We also computed the cumulative cross-section for close approach
during the encounter between two identical circular equal-mass binaries
and compared them with those of Bacon, Sigurdsson \& Davies (1996) 
and Cheung et al. (in preparation).
The comparison with the most recent results by Bacon et al. and
with the results by Cheung et al. indicates that the cross-sections 
are consistent within the numerical uncertainties.

To test our 4-body scattering package against the 3-body package
we reproduced the results by Rasio, McMillan and Hut (1995)
regarding the formation of the triple system PSR B1620-26 in M4.
Branching ratios and cross-sections obtained simulating
binary-binary encounters are well consistent with the ones
obtained in the case binary-single star encounters.

\section{Numerical simulations}
\label{sec:simulations}
\subsection{Initial conditions}
\label{sec:initial}
We consider interactions between a target binary composed by $\aea$ 
and the primary component of $\ioo$ ($\oria$) 
and a projectile  binary composed by $\muc$ and the secondary component 
of $\ioo$ ($\orib$). 
The values of masses, radii and ages used as initial conditions for the simulations
are reported in table \ref{tab:scheme}. 
After the two binaries and their relative orbits are selected we integrate the
equations of motion until the encounter is resolved (see appendix \ref{app:stop}).
The relative velocity at infinity between the centers of mass 
of the two binaries is set equal to the mean dispersion velocity 
in the Trapezium cluster (2$\kms$, Herbig \& Tendrup 1986).
Conservation of total energy and momentum in the 4-body
system center of mass frame is applied to derive the binding
energies of the initial binaries. 
Only the sum of the two binding energies can be derived from the conservation laws,
so the ratio between the two remains as a free parameter:
$\alpha=E_{\rm t}/E_{\rm p}$.

\begin{figure*}
  \caption{Left: branching ratios for different types of outcome as a function of
    the ratio $\alpha$ between the initial binding energies of the two binaries.
    Only flybys, exchanges, ionizations, triples and encounters leading to systems 
    like $\aea$, $\muc$ and $\ioo$ are shown.
    Right: normalized total cross-section as a function of the parameter $\alpha$.}
  \label{fig:branch_sec}
  \begin{center}
    \leavevmode
    \includegraphics[width=8cm]{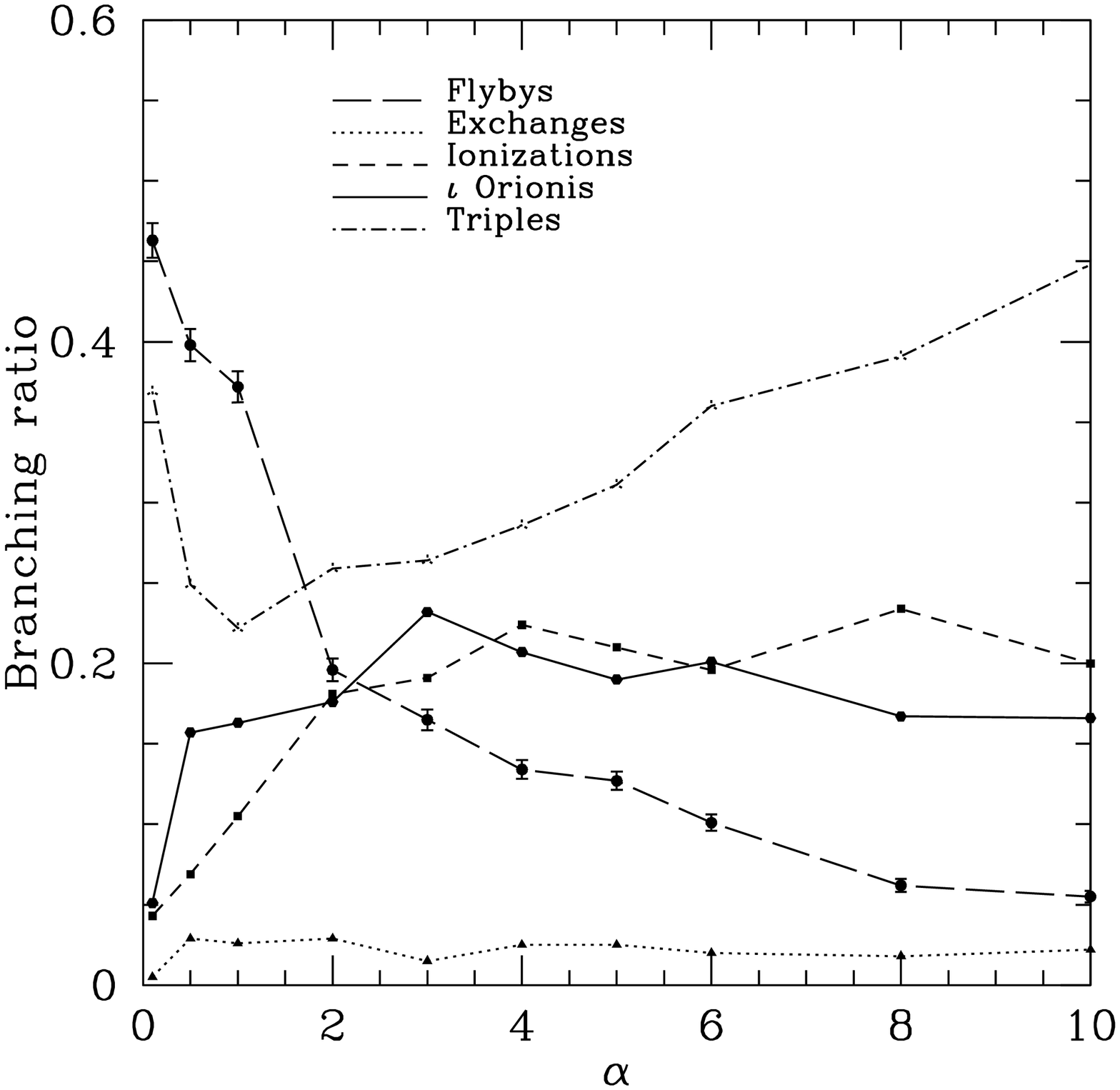}
    \includegraphics[width=8cm]{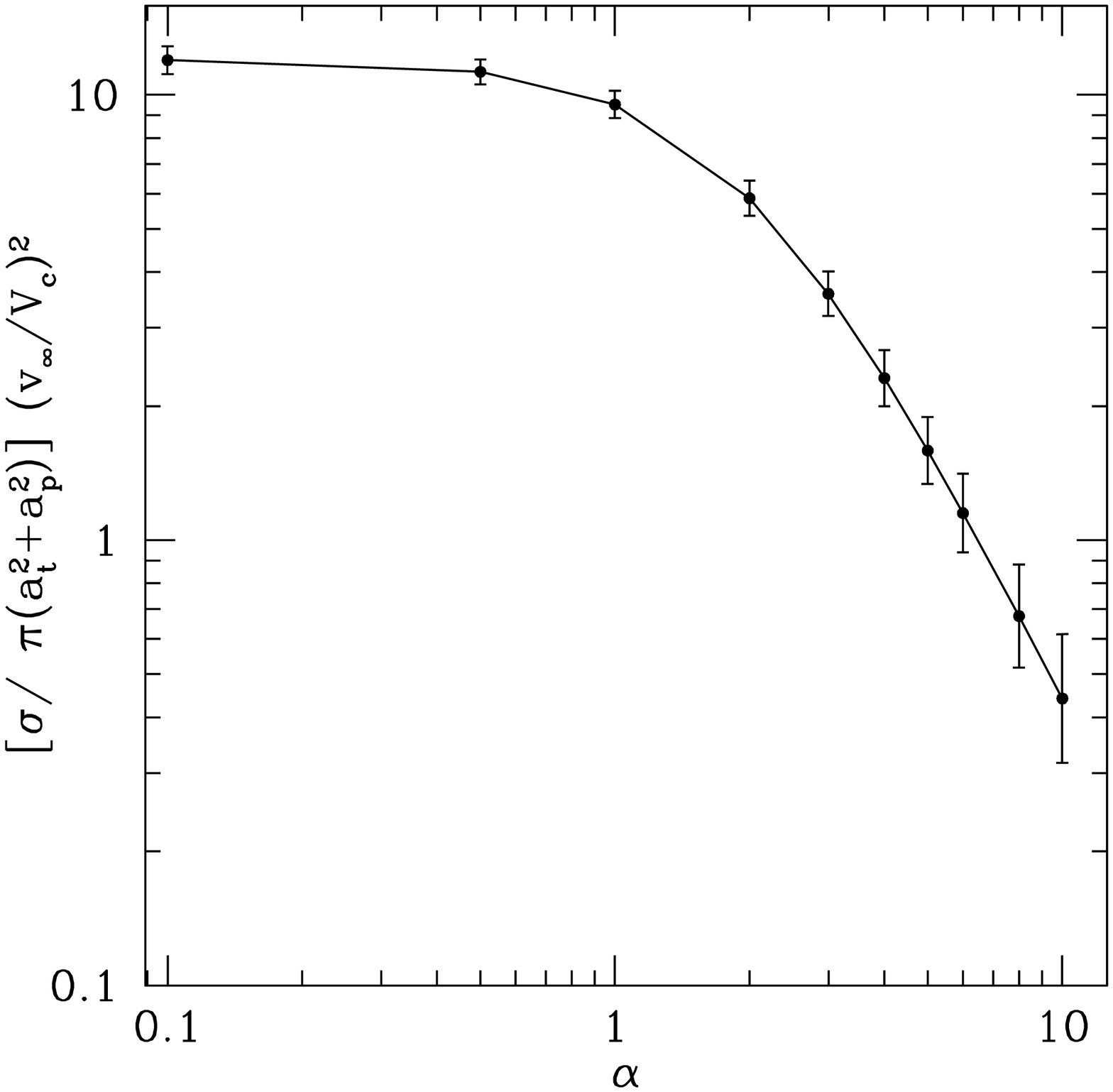}
  \end{center}
\end{figure*}

For each value of $\alpha$ between 0.1 and 10 we perform 2000
scattering experiments and the resulting branching ratios 
and cross-sections are presented in Figure \ref{fig:branch_sec}. 
Most of the encounters lead to the disruption of one binary, 
possibly accompanied by an exchange or a capture in a stable triple.
Only about 10 per cent of the encounters result in an exchange. 
The fraction of flybys decreases as a function of $\alpha$
while the fraction of triples increases for $\alpha>1$. 
A value $\alpha>1$ means that the projectile binary 
is softer than the target binary and is more easily ionized.
The larger is $\alpha$ the wider is the projectile binary. 
After an ionization, one of the component stars can be captured 
by the target binary which contains the most massive star of the system.
 The normalized cross section decreases steadily as a function of $\alpha$
because of the decrease in the normalization factor.
For values $\alpha \simgreat 4$ the cross-section scales as $\alpha^{-2}$.

We mainly focus on encounters of the type
{\scriptsize{$$(\oria,\aea) + (\orib,\muc) \rightarrow (\oria,\orib)+ \aea + \muc$$}}
leading to the ionization of one binary 
and the exchange of one specific star.
This interaction is most favored for $\alpha \simless 3$ 
(see Figure \ref{fig:branch_sec}). 
For clarity we select $\alpha$=2 and for this value
we perform 15000 scattering experiments.

\subsection{The point particle limit}
\label{sec:point}
\begin{figure}
  \caption{Distribution of velocities relative to the center of mass
    of the 4-body system for $\ioo$, $\aea$ and $\muc$ (solid line)
    compared with the observed values (dashed line) for $\alpha$=2.} 
  \label{fig:vel}
  \begin{center}
    \includegraphics[width=7cm]{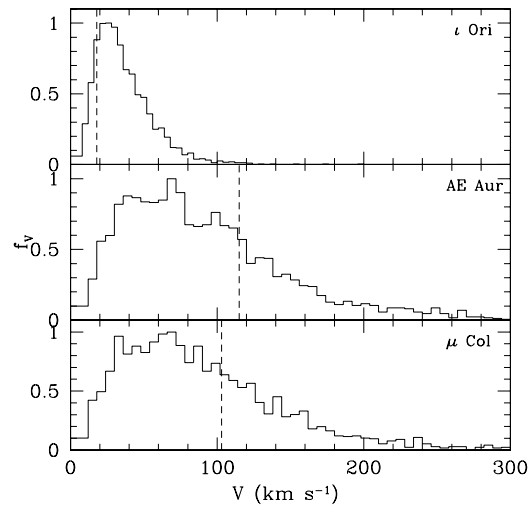}
  \end{center}
\end{figure}

We first analyse the properties of the stars resulting from the
simulations in the approximate case of point particles.
The final velocity distributions for $\ioo$, $\aea$ and $\muc$ 
are shown in Figure \ref{fig:vel}. The velocities are relative 
to the center of mass of the 4-body system and the observed values 
(see Table \ref{tab:vel}) are presented as dashed lines.
Due to momentum conservation the two single stars recoil 
with a velocity $\ts3$ times higher than the binary.
Typical velocities for the stars range from 30 to 100$\kms$,
with an average of the order of their orbital velocities ($\ts 60-70\kms$)
in the initial binaries.
The velocity of the binary has a peak at about 25$\kms$, close to
the observed velocity $18\th\pm\th1\kms$.

\begin{figure*}
\caption{Velocity of $\aea$ versus velocity of $\ioo$ (left) 
and versus velocity of $\muc$ (right). 
The density of points over squared areas is indicated through
a gray shaded scale. The error boxes of the measured velocities 
are shown as filled black rectangles.}
\label{fig:crossv}
\begin{center}
\includegraphics[width=8cm]{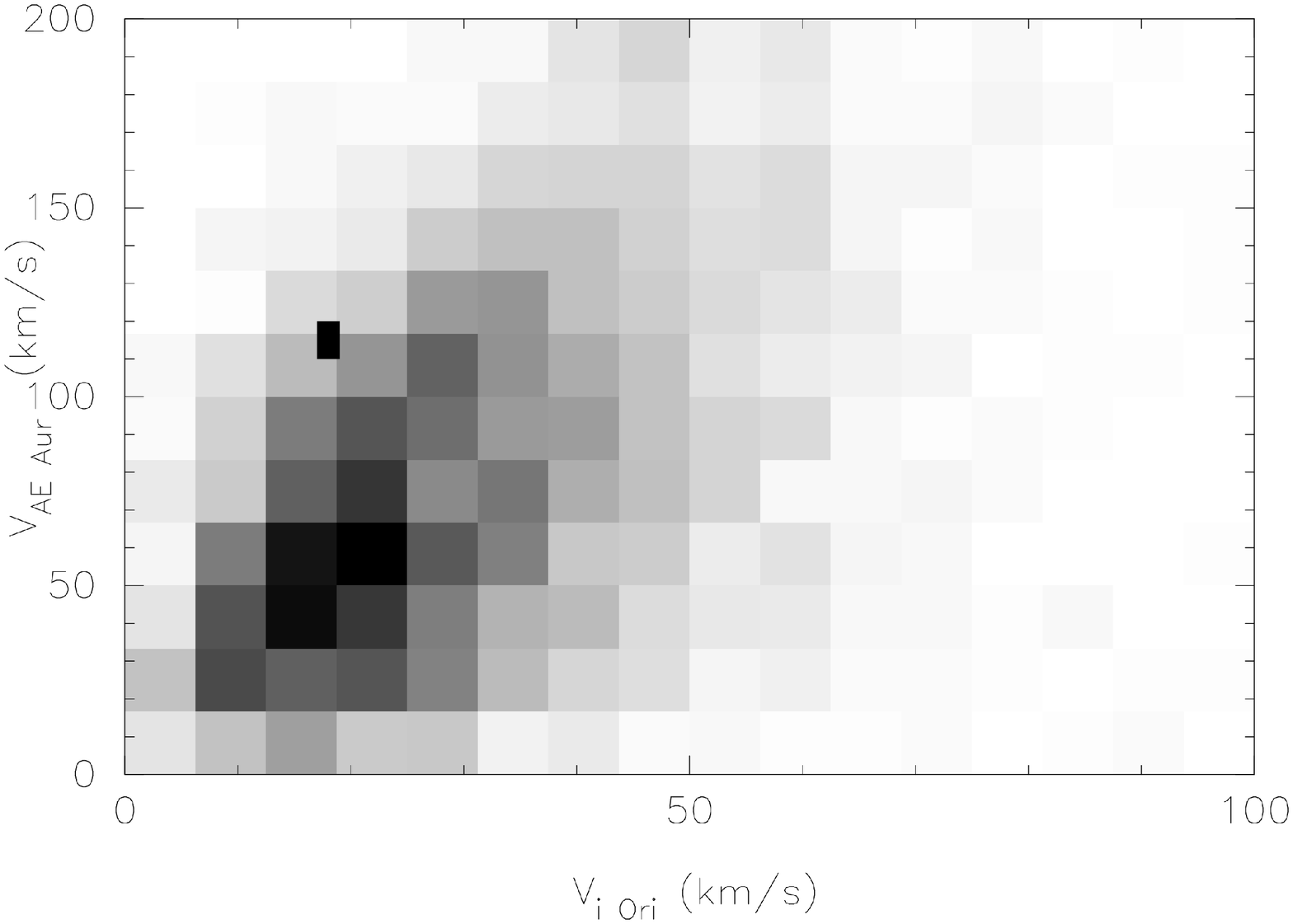}
\includegraphics[width=8cm]{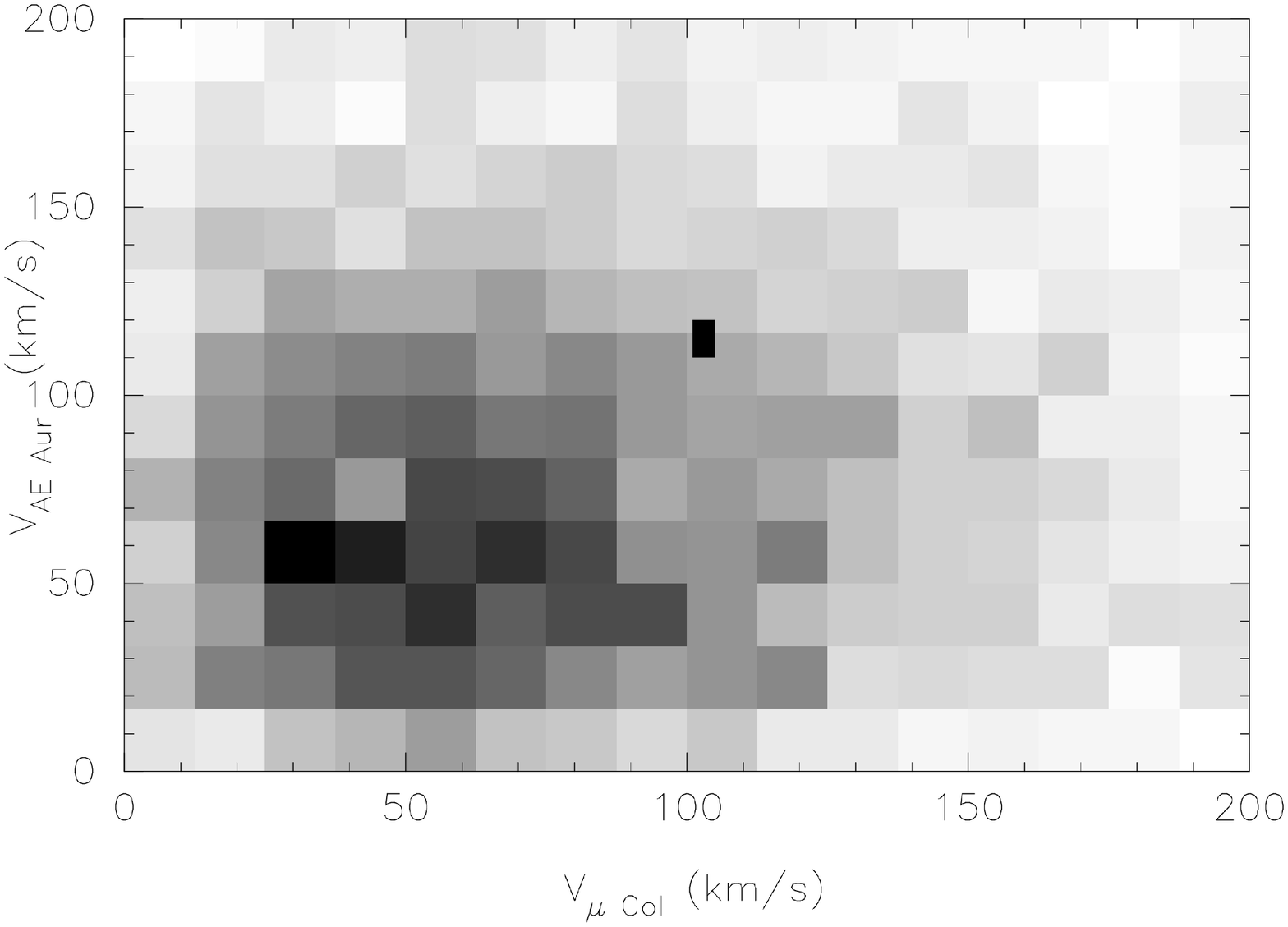}
\end{center}
\end{figure*}

We observe a correlation between the velocity of the binary
and that of the single stars, though with a large scatter,
while there is no apparent correlation between the space velocities 
of the two single stars (see Figure \ref{fig:crossv}). 
This might be due to the many degrees of freedom present in 4-body scatterings.

\begin{figure}
  \caption{Distribution of the final binary semi-major axis (solid line) 
    compared with the observed value (dashed line) for $\alpha$=2.}
  \label{fig:semi}
  \begin{center}
    \includegraphics[width=7cm]{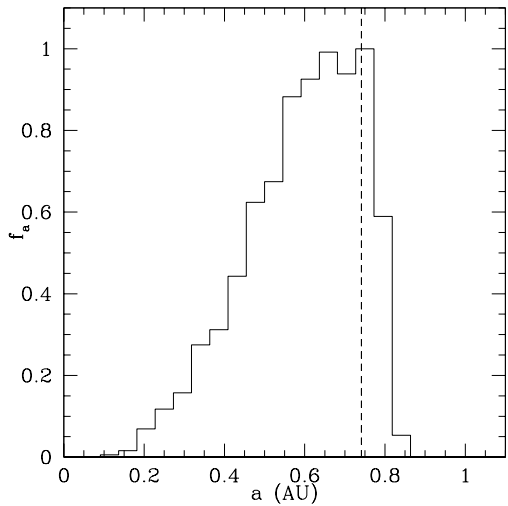}
  \end{center}
\end{figure}

In Figure \ref{fig:semi} we show the distribution of the semi-major axis
of the binary after the encounter. 
The observed value (Stickland 1987; Marchenko et al. 2000)
is consistent with the most probable value in the distribution. 

The distribution of velocities and semi-major axis
does not change for $\alpha \simless 3$ but become significantly
different for larger values of $\alpha$.
The consistency of the distributions with the observed values
reflects the right choice of the total energy in the encounter.
The total energy of the 4-body system is known at present
and its conservation before and after the encounter was used 
in the choice of the initial conditions. Different choices
of the total energy available in the interaction lead to different
distributions for the velocities and the binary semi-major axis. 
This seems to exclude the possibility that a fifth body
was involved in the encounter. 

\begin{figure}
  \caption{Distribution of the relative angle $\theta$ between the 
    velocity vectors of $\aea$ and $\muc$ (solid line) with
    respect to the center of mass of the 4-body system.
    The dashed line indicates the value of the relative angle 
    derived from observations.}
  \label{fig:teta}
  \begin{center}
    \includegraphics[width=7cm]{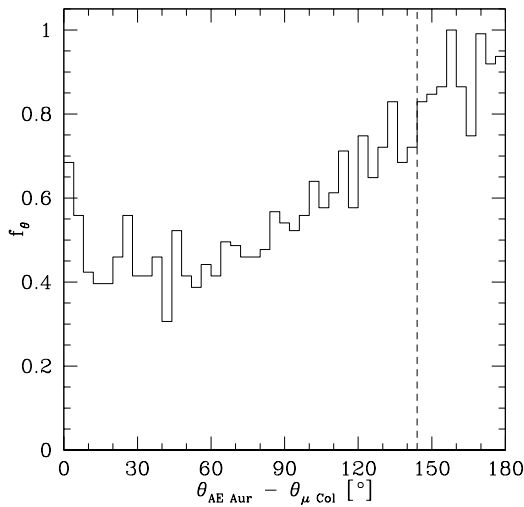}
  \end{center}
\end{figure}

In order to verify the possibility of producing two single runaways
moving in almost opposite directions, as is observed for $\aea$ and $\muc$,
we study the distribution of the angular displacement of the two stars
with respect to the center of mass of the 4-body system. 
We compute the angle $\theta$ between the velocity vectors
of the two single stars and report its distribution $f_{\theta}$ 
in Figure \ref{fig:teta}. We find that about 35 \% of all 
encounters result in an angular displacement $\theta$ in the 
45 degrees range $135^{\circ} \leq \theta < 180^{\circ}$.

The eccentricity of the binary after the encounter is consistent
with a thermal distribution and therefore more than
50 per cent of the encounters result in a binary 
with $e>0.67$.

We computed the absolute fractions and the cross-sections of different types 
of encounters using the procedure described in appendix \ref{app:cross}.
If $\sigma_X$ is the cross-section for process X, 
a normalized cross-section can be defined as
\begin{equation}
  \tilde \sigma_X \equiv \frac{\sigma_X}{\pi~\left(\atar^2+\apro^2\right)}
  \left(\frac{v_{\infty}}{V_{\rm c}}\right)^2,
\end{equation}
where $v_{\infty}$ represents the relative velocity between the centers
of mass of the interacting binaries.
As shown in Table \ref{tab:sec}, about 1 in 6 encounters results in 
the ionization of one binary and the exchange of one specific star, 
leading to systems like $\aea$, $\muc$ and $\ioo$.

From the cross-section it is possible to derive the typical time-scale
of the process under consideration if the stellar density and the stellar 
velocity dispersion of the region are known
\begin{equation}
  T_X \ts \frac{1}{\pi \left(\atar^2+\apro^2\right)} \frac{v_{\infty}}{V_{\rm c}^2} 
  \frac{1}{n~f_b~\tilde \sigma_X}, 
\end{equation}
where $n$ represents the mean stellar density in the cluster 
and $f_b$ is the fraction of binaries in the core.
Substituting typical parameters of the Trapezium cluster 
(with a binary fraction $f_b \ts 0.6$; Duquennoy \& Mayor 1991) 
we obtain
\begin{equation}
  T_X \ts \frac{3\th{\rm Myr}}{\tilde \sigma_X} \left(\frac{5\times10^4\th\pct}{n}\right) 
  \left(\frac{\rm AU^2}{a_t^2+a_p^2}\right) \left(\frac{v}{0.01}\right)^2 
  \left(\frac{5 \kms}{v_{\infty}}\right)
\end{equation}
where $v=v_{\infty}/V_{\rm c}$. 
In the case $\alpha$$\th=\th$2 the typical time-scale for a binary-binary encounter
resulting in an exchange-ionization is about 3$\th$Myr.

\begin{table}
  \caption{Branching ratios $f_X$ and normalized cross-sections 
    $\tilde \sigma_X$ for different types of outcomes in the case $\alpha$=2.  
    $1\sigma$ error bars are of the order of 1 per cent.  
    We specifically split off encounters which lead to a system 
    like $\ioo$, at the bottom, below the horizontal line.
  }
  \label{tab:sec}
  \begin{center}
    \begin{tabular}{lcc}
      \hline
      Encounter           & $f_X$   & $\tilde \sigma_X$ \\
      \hline
      \hline
      Preservation        &  0.21   &   1.20\\
      Exchange            &  0.03   &   0.18\\
      Ionization          &  0.17   &   1.00\\
      Exchange-ionization &  0.35   &   2.06\\
      Triple              &  0.24   &   1.42\\
      Total ionization    &  0.00   &   0.00\\
      \hline
      $\ori$              &  0.20   &   1.17\\
      \hline
    \end{tabular}		
  \end{center}
\end{table}

\subsection{Collisions in binary-binary encounters}
\label{sec:collisions}	

In this section we relax the hypothesis of point mass stars and
investigate the effects of physical radii on the outcomes of
binary-binary encounters.  In the simulations described in the 
previous sections, we keep track of the minimum distances 
between all stars. In this way we are able to use
the same data and test the effect of finite stellar radii, assuming
that stars collide if they approach each other within some minimum
distance $d_{\rm coll}$.

Since the initial eccentricities of both binaries are randomly drawn
from a thermal distribution between 0 and 1, stars in very eccentric
orbits may collide at the first pericenter passage.  This would not be
realistic, and therefore we limit the range in initial eccentricities.
In a highly eccentric orbit two stars with total radius $R$ collide if
$R \simgreat 0.75 a (1-e)$ (Kopal 1959).  Inverting this equation
leads to our definition of $\emax$.   For the observed stellar radii
(see Table \ref{tab:scheme}) $\emax= 0.9$. We therefore exclude
binaries with initial eccentricity $e > 0.9$ from further
consideration.

\begin{figure}
  \caption{Cumulative distribution for minimum distances $r_{\rm min}$, 
    between stars during binary-binary encounters.  
    The solid line represents the result for systems with $e < 0.9$ 
    from our standard run.  The dashed line is for binaries with 
    initially circular orbits.  With the observed stellar radii a 
    collision would occur at about $r_{\rm min}/a \simless 0.07$.}
  \label{fig:cumulative} 
  \begin{center}
    \includegraphics[width=7cm]{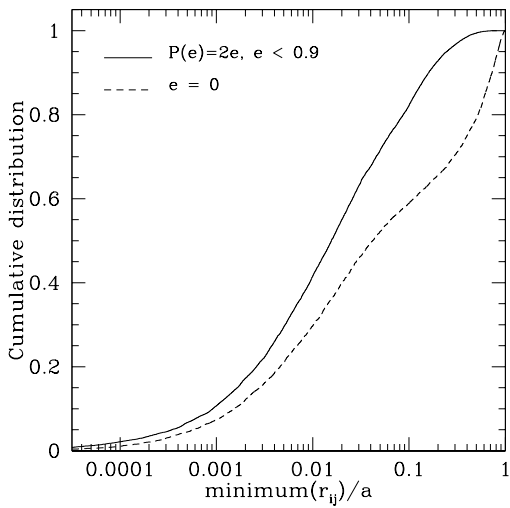} \end{center}
\end{figure}

\begin{figure}
  \caption{Fraction of collisions as a function of $\emax$.}
  \label{fig:collemax}
  \begin{center}
    \includegraphics[width=7cm]{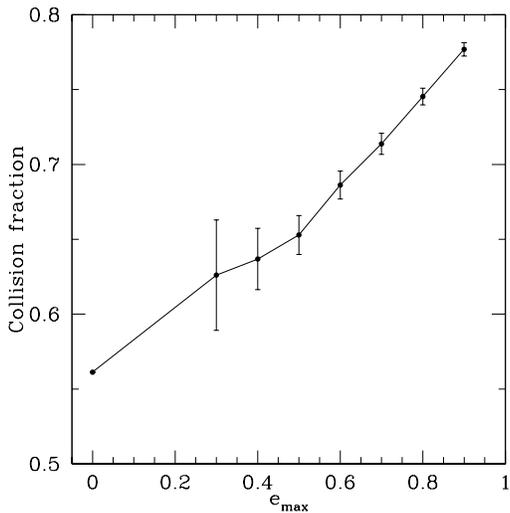}
  \end{center}
\end{figure}

Figure \ref{fig:cumulative} shows the cumulative distribution for
minimum distances between any two stars for the $\ts 8000$ encounters
which remained after selecting the systems with $e < \emax$ (solid curve) 
and for circular initial binaries (dashed curve).  
Since circular orbits are very rare in our standard initial conditions, 
we performed a separate simulation of 8000 encounters. 
If we assume stellar radii from Tab.\,\ref{tab:scheme} at the moment of the
encounter, a collision occurs if $r_{\rm min}/a \simless 0.07$. 
In that case about half of the initially circular binaries result in a
collision, whereas only 25 per cent of the eccentric systems survive
merging.  In figure \ref{fig:collemax} we show the fraction of
binaries that experience a collision as a function of $\emax$.  
The two figures \ref{fig:cumulative} and \ref{fig:collemax} show similar
trends; high initial eccentricities enhance the collision rate.

\begin{figure}
  \caption{Branching ratios as a function of $\emax$. 
    The solid line gives the results for encounters in which one binary 
    was ionized.  The dashed line gives results for a subset of these, 
    namely the ones which lead to systems like $\ioo$ (see $\S$ \ref{sec:initial} 
    for the specifics of this encounter).}
  \label{fig:branchemax}
  \begin{center} 
    \includegraphics[width=7cm]{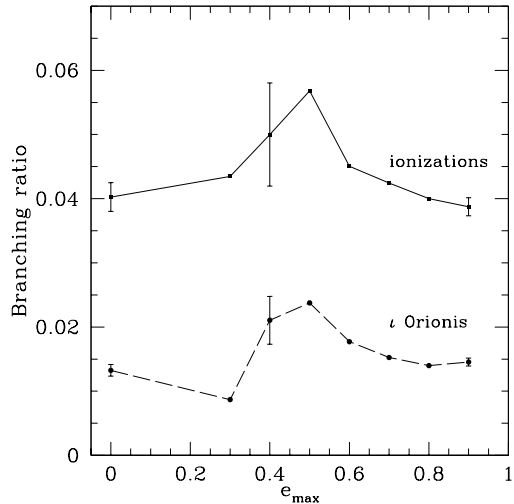}
  \end{center}
\end{figure}

In Figure \ref{fig:branchemax} we show the proportion of ionized
systems and systems like $\ioo$ which resulted from the standard run.  
There seems to be a slight preference for initial eccentricities 
in the range $0.4 \simless e \simless 0.6$.
 
\begin{figure}
  \caption{The fraction of collisions (dotted line) and the branching
    ratio for ionizations (solid curve) as a function of $\beta$.  Our
    system of interest is represented with the dashed line.  }
  \label{fig:beta} 
  \begin{center}
    \includegraphics[width=7cm]{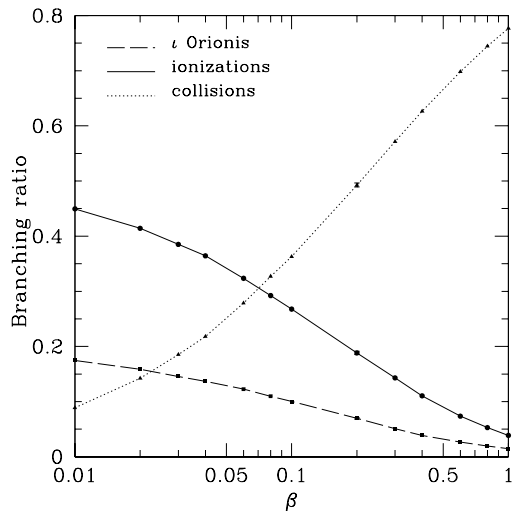} \end{center}
\end{figure}

To further study the dependence of the collision probability on the
stellar radii we define $\beta \equiv d_{\rm coll}/r$, the dimensionless
effective radius for collisions.  In Figure \ref{fig:beta} we show the
branching ratios for collisions, ionizations and the sub-type of
$\ioo$ as a function of $\beta$.  Changing the stellar radii (or
$\beta$) by a factor of a few does not effect the collision rate much,
and therefore the choice of the stellar radii is not critical to the
results.

\section{The evolution of $\iota$ Orionis after the encounter}
\label{sec:evolution}
Until now we have considered the implications of a binary-binary 
encounter on the dynamical properties of $\aea$, $\muc$ and $\ioo$. 
The current observations of the binary allow us to constrain 
its evolution in the $\ts2.5\th$Myr elapsed from the moment 
of the encounter till now.
There are at least three further aspects that have to be integrated 
into a wholly plausible model of the interaction under consideration. 
They are:
\begin{enumerate}
\item $\ioo$ might be a triple system, and possibly even 
  quadruple or quintuple (see $\S$ \ref{sec:speckle});
\item stellar evolution models make it difficult to confirm
  that $\orib$ is of the same age as any of the other three stars 
  (see $\S$ \ref{sec:age});
\item tidal friction must operate within the present $\ioo$ system, 
  and we have to check whether the observed eccentricity 
  can have been maintained for the 2.5$\th$Myr since the collision 
  (see $\S$ \ref{sec:tidal}).
\end{enumerate}	
We deal with these in turn.

\subsection{Is $\iota$ Orionis a multiple system?}
\label{sec:speckle}
The spectroscopic binary $\ioo$ is the brightest (component A) of three members 
of the multiple star ADS 4193. The other two members (B and C) are 
$11\asec$ and $50\asec$ distant. At the distance of $\ioo$ 
($\ts 400\th$pc, see Table \ref{tab:vel}) these are quite widely separate, 
and for the time being we discount them as gravitationally bound companions. 
But ADS 4193A is also a speckle binary, CHARA 250Aa/Ab,
with a separation of $0.10\asec$ (Mason et al. 1998). 
This suggests an orbital period $\simgreat 40\th$yr. 
The brighter component (Aa) of this speckle pair is the spectroscopic
binary. Strictly speaking we should refer to the O9III and B0.8III-IV components of this
$29\th$d binary as $\oriaao$ and $\oriaat$, and we do so from now on. 

The close speckle companion raises at least two questions. Is it reasonable that the
companion was carried along during the four-body (or rather 5-body) encounter that
we have supposed? And could it affect the eccentricity of the spectroscopic orbit, by the 
mechanism of Kozai cycles (Kozai 1962)? 

We have not attempted to model 5-body encounters because of the enormous 
parameter space that they can be drawn from. We can imagine
it would be possible for the companion to remain bound 
after an encounter but we doubt the energy released
would be enough to eject $\aea$ and $\muc$ with high speed. 
It might be worth investigating this possibility in the future
and testing these predictions.

In a triple system in which the outer orbit is fairly highly inclined 
to the inner orbit, the third body can have a major long-term effect 
on the inner pair, causing its eccentricity to cycle between a low 
and a high value. For instance, if the mutual inclination is $60^\circ$, 
$e$ can cycle between 0 and 0.76, or between 0.5 and 0.86 
(Kiseleva, Eggleton \& Mikkola 1998). The period of the cycle is 
$\ts P_{\rm out}^2/P_{\rm in}\ \ts\ 20000\th$yr in this case. 
If the triple were formed early in the life of the cluster by some random encounter, 
an inclination of $60^\circ$would be quite likely; alternatively, 
the kind of encounter that we think is needed to create
the binary $\oriaa$ would probably randomize its inclination relative to the ($\oriaa, \oriab$) 
speckle pair. Thus it is possible that the observed high eccentricity of the 
spectroscopic binary $\oriaa$ is not a result of the encounter, but of the third body $\oriab$.

\subsection{Stellar evolution models for AE Aurigae, $\mu$ Columbae and $\iota$ Orionis}
\label{sec:age}
\begin{figure}
  \caption{Evolutionary tracks in the HR diagram for log $M$=1.1, 1.125,...,1.5.
    Also shown (thick lines) are isochrones for ages 5 Myr and 10 Myr.
    Sloping nearly-straight dotted lines are for log $g$=2.8, 3.0,...,3.8,
    starting from the upper right corner and ending near the ZAMS. 
    Approximate positions for $\oriaao$, $\oriaat$, $\aea$ and $\muc$,
    based on their temperatures and gravities, are indicated 
    by Aa1, Aa2, AE and $\muc$ respectively. In the scheme the epoch of 
    formation of the eldest binary is labelled as $T=0$, the moment
    of the encounter as $T=4.5\th$Myr and the present
    epoch as $T=7\th$Myr.}
  \label{fig:hr}
  \begin{center}
    \includegraphics[width=8.0cm]{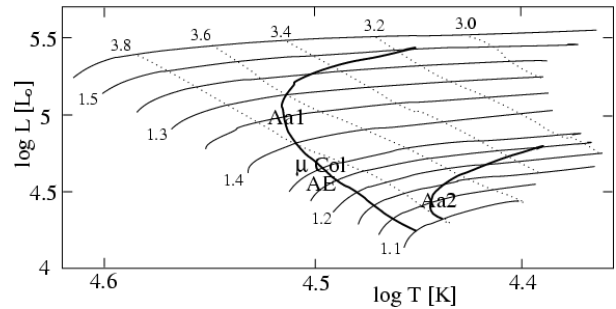}
  \end{center}
\end{figure}

We now use stellar evolution model to constrain the ages of the four stars.
Figure \ref{fig:hr} is a theoretical HR diagram for masses in the range required. 
The positions of all four stars, according to their effective temperatures 
and gravities as given by Bagnuolo et al. (2001), are indicated. 
Two isochrones are indicated, for 5 and 10 Myr. It is clear that the apparent ages of 
$\oriaao$ and $\oriaat$ are substantially different: we estimate
5 and 10 Myr respectively. The uncertainties appear to be of the order of $\pm 1\th$Myr,
but could be substantially greater on account of unquantifiable systematic errors.
The possibility that discrepant ages can be accounted for by Roche-lobe
overflow can almost certainly be discounted by the high eccentricity of the system.
The ages of $\aea$ and $\muc$ are comparable to that of $\oriaao$, 
though $\aea$ might be a little younger and $\muc$ a little older. 

The Trapezium Cluster, from which the four stars appear to have been ejected, 
is a well studied young cluster, the core of the Orion Nebula Cluster, 
which contains about a dozen OB stars and many more fainter stars. 
It is estimated that these stars are $\simless 1\th$Myr old 
(Herbig \& Terndrup 1986; Brown, de Geus \& de Zeeuw 1994; Prosser et al. 1994). 
Stars later than B2 are often though not always above the main sequence. 
We may therefore wonder if the two giants of $\ioo$\ are still evolving 
{\it towards} the main sequence rather than away from it. 
This seems improbable, because their ejection from the cluster $2.5\th$Myr 
ago is amply long enough for contraction to the main sequence at their high 
luminosities. This suggests, incidentally, that not {\it all} the Trapezium stars 
are as young as $1\th$Myr, and we appear to require that some massive stars have 
been forming over the last $10\th$Myr. An age spread from 1 to 10$\th$Myr has already
been observed for stars $\simless 6\msun$ (Palla \& Stahler 1999), and we suggest that 
there may be a similar spread for OB stars. 

An alternative possibility is that the parent cluster is one of the older
subgroups in Orion such as subgroup 1a or 1c (see Brown et al. 1998).
These subgroups have a kinematics very similar to the one of subgroup 1d 
(which contains the Trapezium cluster) and overlap in position on the sky with $\ioo$. 

The uncertainties in the theoretical HR diagram are of course substantial, 
and hard to quantify. The models were computed using the code of Pols et al. (1997), 
which was found to give good agreement with the observational data of Andersen (1991).
The theoretical uncertainties have to be convolved with observational uncertainties
but very substantial error must be involved if $\oriaat$ is to be of the same 
age as any of the other three stars.

A rather long shot is the possibility that the speckle companion is in fact 
the missing $10\th$Myr-old companion of $\orib$. 
This would require a binary-triple interaction of the form:
$$(\oriaat , \oriab) + (\oriaao , (\aea , \muc)) \rightarrow$$
$$((\oriaao , \oriaat) , \oriab) + \aea + \muc.$$
It is hard to assess the likelihood of this but in any event we should not 
ignore the possible existence of a fifth body in a complete analysis.

\begin{figure*}
  \caption{The evolution with age of orbital and spin parameters subject to tidal
    friction. The present epoch is at $2.5\th$Myr. (a) Left panel: eccentricity 
    (decreasing), and the cosine of the inclination $\zeta$ of the stellar spin axis to the 
    orbital axis (increasing; $\oriaao$ only). The components were started with rotational 
    axes oblique to the orbital axis; nutation on a short time-scale causes the apparent 
    breadth of the inclination curve before parallelization is complete. Interference
    between the nutation frequency and the rather low data-sampling frequency causes some
    artefacts in the curve. (b) Right panel: orbital period (upper curve), and spin 
    period (lower curve; $\oriaao$ only).}
  \label{fig:tidal}
  \begin{center}
    \includegraphics{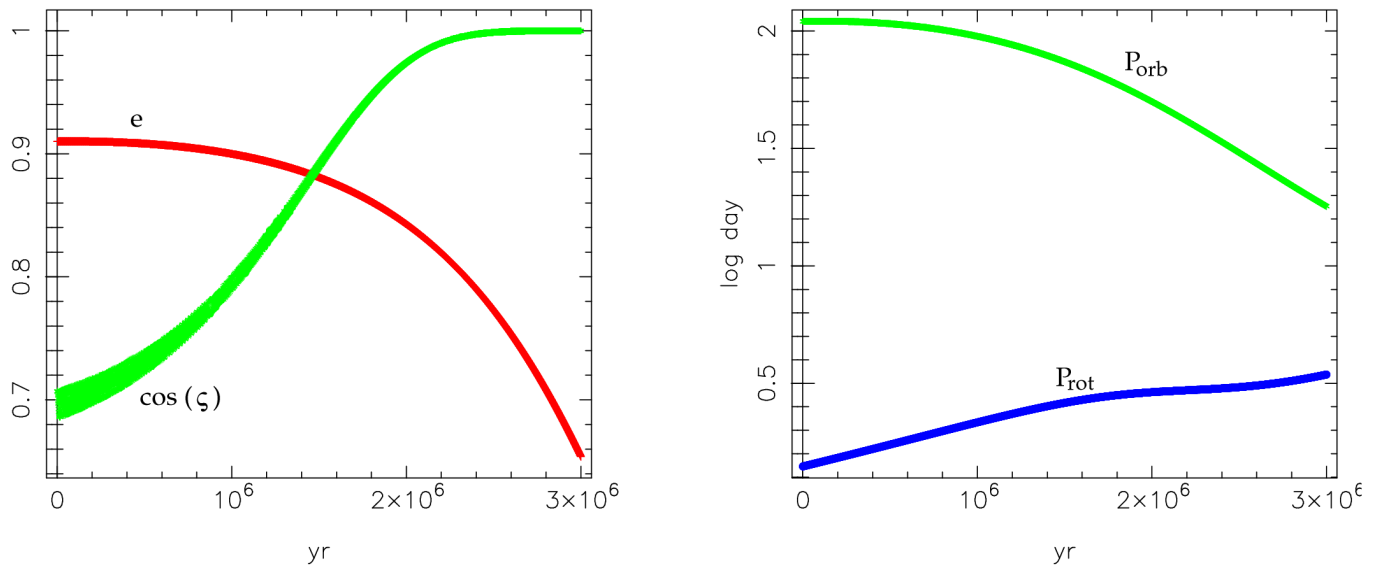}
  \end{center}
\end{figure*}

\subsection{The effects of tidal friction on the binary orbit}
\label{sec:tidal}
An important aspect to consider in the post-encounter evolution
is the effect of tidal friction on the spectroscopic orbit.
The sum of the radii that we infer is $60\%$ of the periastron separation, which
would seem sufficient for tidal effects to be important. We have used the 
equilibrium-tide model of Hut (1981) as further developed by Eggleton, Kiseleva \& 
Hut (1998). This model was successfully tested on the SMC radio-pulsar 
binary 0045-7319 (Eggleton \& Kiseleva (2001)).

The orbital evolution expected since the encounter 2.5$\th$Myr ago 
is shown in Figure \ref{fig:tidal}. We find that we could reach 
the present eccentricity and orbital period if we started with 
$P$=110 days and $e$=0.91. 
A high eccentricity is common for binaries which undergo exchange-ionization
encounters as the final eccentricity distribution is expected to be thermal
(see $\S$ \ref{sec:point}).
On the other hand, a long-period binary as a final outcome implies
an even wider initial target binary, which means a lower binding energy
available in the encounter. It is unlikely to obtain
recoil velocities as high as 100$\kms$ with a pre-encounter total energy
which is smaller than the current one. 
A tight binary (with an orbital period similar to the observed one) 
is needed in order to reproduce the velocities of the single runways. 

Figure \ref{fig:tidal} shows that parallelization (the alignment of the spin axis
of the two binary components) and pseudo-synchronization 
took $\ts 2\th$Myr to achieve. Given the uncertainty in the viscous time-scale 
-- at least a factor of 10 -- it is possible that they have actually not yet 
been achieved. The pseudo-synchronous rotation period (Hut 1981) 
is a fraction of the orbital period dependent only on eccentricity: 
the factor is 9.51 for $e$=0.76, i.e. $P_{\rm rot}\ts 3\th$d currently.
With the radii computed from the evolutionary tracks of Figure \ref{fig:hr}, the 
rotation speeds we expect are $V\sin i \ts 150$ and 115$\th\kms$
for the primary and the secondary respectively. 
These values are somewhat greater than the values (110 and 70$\th\kms$) 
quoted by Marchenko et al. (2000), and may indicate that 
pseudo-synchronism has not yet been reached. 

However we have already noted that the high eccentricity of the spectroscopic
binary may not in fact have been generated by the collision but may be an artifact
of Kozai cycles driven by the third (speckle) body (see $\S$ \ref{sec:speckle}). 
In Kozai cycles, while the eccentricity fluctuates the semimajor axis 
is unperturbed (to lowest order), and as a result the two components 
will in the long term seldom be as close together as they are at 
the periastra of the current orbit. 
Thus tidal friction might be negligible in this case.

The possibility that the eccentricity is due to the third body rather
than to the interaction could mean that we have to revise our earlier conclusion
that the apparently non-coeval nature of the spectroscopic binary is not due
to Roche lobe overflow, since the main argument against there having been 
any Roche lobe overflow is that the orbit is eccentric. 
However Kozai cycles are normally suppressed if the
quadrupolar distortion of the stars in the close binary is large; and it
usually {\it is} large enough as a star approaches Roche lobe overflow. 
Once a star fills its Roche lobe it normally grows larger still, 
until a very late stage of evolution when it has lost $\ts 70\%$ of its mass. 
Thus once Kozai cycles are suppressed they are unlikely to re-establish 
themselves till this late stage, when the donor would be a small helium star 
rather than a B giant.

We hope that in the near future more information will be available on
the speckle orbit, which should move appreciably in $\ts 10\th$yr. The
spectroscopic orbit may also be on the margin of detectability by speckle
or adaptive optics; a direct measure of its inclination could limit the
possibilities substantially. It will be desirable to search for the speckle
companion in the spectrum of $\ioo$. If it can be detected it may modify
significantly the spectroscopic elements of the 29$\th$d orbit.

\section{Summary}
\label{sec:disc}

We tried to reconstruct the event that ejected the runaways
$\aea$ and $\muc$ and the binary $\ioo$ from their parent cluster
about 2.5$\th$Myr ago. We here summarize different models that could 
explain the high velocities of the single stars
and the discrepant ages of the binary components.

\begin{itemize}
\item [1.] The single stars and the binary are all unrelated.
As shown by Gies and Bolton (1986) and Hoogerwerf et al. (2000, 2001), 
this can be excluded on the basis of their kinematical and evolutionary properties.  
The same dynamical encounter must have ejected the stars from the Trapezium about 2.5 $\th$Myr ago. 
In order to explain the high velocities of the single stars and the age difference in
the binary, the encounter must have involved an ionization and an exchange.  

\item [2.] The stars were ejected as a result of a binary-binary interaction
that occurred in the Trapezium cluster about 2.5$\th$Myr ago.
This model is based on the observation that the velocity of the
4-body system is comparable to the velocity of the Trapezium cluster.
The velocities of the stars and the period of the binary
after the encounter are consistent with the observed values, 
so that we can conclude that this model can well reproduce 
the kinematics of the 4-body system.
If we consider stars with physical radii, a fraction 
of encounters result in a collision between two stars. 
We thus expect to find collision products in dynamically active 
young clusters like the Trapezium.
The fraction of collisions strongly depends on the initial eccentricity
of the two binaries and therefore we consider it likely
that $\ioo$, $\aea$ and $\muc$ were ejected in
an encounter between two low-eccentricity binaries.
In this case, the final properties of the system like the recoil velocities
of the stars and the orbital period and eccentricity of $\ioo$ do not differ 
from the ones obtained for highly eccentric initial binaries.
The occurrence rate of this type of encounter, once corrected
for the collision probability, is about one in 10 Myr.

A possible difficulty in this model is the short circularization
time-scale for the orbit of the binary after the encounter.
Our calculations show that only extremely eccentric 
post-encounter orbits ($e \ts 0.9)$ evolve in an eccentricity 
like that of $\ioo$ after 2.5 Myr. 

In addition, the model relies on the assumption that $\orib$ and $\muc$ 
are in the same binary before the encounter, 
implying that the two stars must be coeval. 
Our evolutionary calculations show that $\muc$ might be too hot,
or equivalently too early in spectral type, to be on the same
isochrone as $\orib$. 
However, our knowledge about the temperatures and gravities 
of the stars is sufficiently imprecise that we can assume
coevality for all four stars.
  
\item [3.] The stars were ejected as a result of a triple-binary interaction.
This model is based on the possibility that $\ioo$ is the brightest
component of a speckle pair, making it a stable hierarchical triple.
This would require a 5-body encounter with the same 
total energy as our standard 4-body encounter 
resulting in the ejection of a triple and two single stars.
According to this model, the high eccentricity of $\ioo$ 
may be a result of Kozai cycles driven by the third body.
We haven't tested this possibility but we can argue that this type 
of encounter would be rather rare
and not energetic enough to explain the velocities of the runaways.
It may certainly be worth investigating this scenario 
to test these predictions as well as the ones about corotation 
and pseudo-synchronisation.
\end{itemize}

The last word has evidently not been said on the interesting interaction
between $\ioo$, $\aea$ and $\muc$, but we think it remains a fascinating 
combination of orbital dynamics, cluster dynamics, stellar evolution 
and tidal effects. 
In addition to more elaborate numerical simulations, high accuracy parameters 
for the stars are needed in order to settle the co-evality issue.
New interferometric speckle observations, 
possibly resolving the spectroscopic binary too, will be crucial
for a better understanding of the dynamical encounter
that ejected the runaway stars.

\section{Acknowledgments}
We would like to thank Douglas Gies, William Bagnuolo, 
Lex Kaper, Anthony Brown, Ronnie Hoogerwerf and Michela Mapelli 
for interesting discussions and comments on the manuscript.
This work was supported by the Netherlands Organization 
for Scientific Research (NWO), the Royal Netherlands Academy 
of Arts and Sciences (KNAW) and the Netherlands Research School 
for Astronomy (NOVA). 
This work was performed under the auspices of the U.S. Department of Energy,
National Nuclear Security Administration by the University of California, 
Lawrence Livermore National Laboratory under contract No. W-7405-Eng-48.

\appendix

\section{Computation of cross-sections}
\label{app:cross}
The procedure described in section $\S$\ref{sec:code} for the 
computation of the maximum impact parameter in each scattering 
experiment is incorporated in the actual computation of cross-sections.
The cross-section for an event X is derived from (McMillan \& Hut 1996)
\begin{equation}
 \sigma_X=\sum_{i=0}^{i_{max}} \pi \left(b_{i+1}^2-b_i^2\right)
\left(\frac{N_{X,i}}{N_i}\right),
\end{equation}
where $N_{X,i}$ is the total number of scatterings resulting in
an outcome X in the shell $i$ and $N_i$ is the total number of scatterings 
performed in the shell $i$.
The uncertainty in the cross-section is given by
\begin{equation}
  \Delta \sigma_X = \sum_{i=0}^{i_{max}} \pi \left(b_{i+1}^2-b_i^2\right) 
  \left(\frac{\sqrt{N_{X,i}}}{N_i}\right).  
\end{equation}

In analogy to the case of binary-single star scattering
(Hut \& Bahcall 1983; Sigurdsson \& Phinney 1993) and in order
to compute physical time-scales for different type
of interactions, we define a cross-section 
\begin{equation}
  \tilde \sigma_X=\frac{\sigma_X}{\pi~\left(a_t^2+a_p^2\right)}
  \left(\frac{v_{\infty}}{V_{\rm c}}\right)^2
\end{equation}
normalized to the geometric cross-section and corrected 
for gravitational focusing.

A characteristic time-scale for a process X, defined as
the mean time between subsequent occurrences of X, 
can be evaluated from (Bacon, Sigurdsson \& Davies 1996)
\begin{equation}
  T_X \sim \frac{1}{\pi \left(a_t^2+a_p^2\right)} \frac{v_{\infty}}{V_{\rm c}^2} 
  \frac{1}{n~f_b~\tilde \sigma_X}
\end{equation}
where $n$ represents the mean stellar density in the cluster 
and $f_b$ is the fraction of binaries in the core.

\section{Criteria for stopping the integration}
\label{app:stop}
During the numerical integration of a binary-binary encounter
the status of the system is analysed after each 20 orbital periods 
of the initial target binary and the simulation is stopped 
if one of the following criteria is met:
\begin{itemize} 
\item [1.] only two stars remain, in which case their orbits are determined 
and the system is classified as a binary if it is bound or as
two single stars if it is unbound;
\item [2.] three stars remain, in which case the system is considered ionized
if the total energy is positive, the stars are widely separated 
and receding from each other while it is considered as a binary and a single star 
if the total energy is negative but the single star is escaping 
from the binary center of mass;
\item [3.] the 4-body system is split in binaries whose centers of mass are unbound,
widely separated and receding.
In the other cases the system is considered bound and the integration 
continues until a stopping criterion is satisfied or a maximum integration 
time is reached.
\end{itemize}

\label{lastpage}

\end{document}